\documentclass[pdflatex,sn-mathphys-num]{sn-jnl}% Math and Physical Sciences Numbered Reference Style
\usepackage{graphicx}%
\usepackage{multirow}%
\usepackage{amsmath,amssymb,amsfonts}%
\usepackage{amsthm}%
\usepackage{mathrsfs}%
\usepackage[title]{appendix}%
\usepackage{xcolor}%
\usepackage{textcomp}%
\usepackage{manyfoot}%
\usepackage{booktabs}%
\usepackage{algorithm}%
\usepackage{algorithmicx}%
\usepackage{algpseudocode}%
\usepackage{listings}%
\theoremstyle{thmstyleone}%
\theoremstyle{thmstyletwo}%

\theoremstyle{thmstylethree}%

\begin{document}

\title[A possible approach to overcome the saturation of the neutron yield in a Plasma Focus and to achieve breakeven]{A possible approach to overcome the saturation of the neutron yield in a Plasma Focus and to achieve breakeven}

%%=============================================================%%
%% GivenName	-> \fnm{Joergen W.}
%% Particle	-> \spfx{van der} -> surname prefix
%% FamilyName	-> \sur{Ploeg}
%% Suffix	-> \sfx{IV}
%% \author*[1,2]{\fnm{Joergen W.} \spfx{van der} \sur{Ploeg} 
%%  \sfx{IV}}\email{iauthor@gmail.com}
%%=============================================================%%

\author*[1]{\fnm{Andrea} \sur{Di Vita}}\email{andrea.divita@ansaldoenergia.com}

%\author[2,3]{\fnm{Second} \sur{Author}}\email{iiauthor@gmail.com}
%\equalcont{These authors contributed equally to this work.}

%\author[1,2]{\fnm{Third} \sur{Author}}\email{iiiauthor@gmail.com}
%\equalcont{These authors contributed equally to this work.}

\affil*[1]{\orgdiv{D.I.C.C.A.}, \orgname{Universita' di Genova}, \orgaddress{\street{via Montallegro 1}, \city{Genova}, \postcode{16145}, 
%\state{State}, 
\country{Italy}}}

%\affil[2]{\orgdiv{Department}, \orgname{Organization}, \orgaddress{\street{Street}, \city{City}, \postcode{10587}, \state{State}, \country{Country}}}

%\affil[3]{\orgdiv{Department}, \orgname{Organization}, \orgaddress{\street{Street}, \city{City}, \postcode{610101}, \state{State}, \country{Country}}}

%%==================================%%
%% Sample for unstructured abstract %%
%%==================================%%

\abstract{Saturation of the neutron yield with increasing energy of the condenser bank in a Plasma Focus ('PF') led to the shutdown of PF research focussed on controlled nuclear fusion in the past. We review available models of saturation and develop further the model of (Lee S., Applied Phys. Lett. \textbf{95}, 151503 (2009)). This model relies on the well-known and generally accepted Lee's model (Lee S., J. Fusion Energy (2014) \textbf{33}:319) of PF discharges and describes saturation in terms of the 'dynamic resistance', i.e. the rate of change of PF inductance due to the motion of the plasma sheath during rundown. A model of this sheath (Di Vita A., J. Plasma Physics (1993), \textbf{50}, 1, pp. 1-19) shows that its spontaneous filamentation rules the dynamic resistance, spoiling the power supply from the condenser bank to the plasma at the values of condenser bank energy ($\geq $ 0.5 MJ) which are relevant to a fusion reactor. Together, these two models lead to the conclusion that suppression of such filamentation prevents saturation, multiplies the PF drive parameter by a factor 3 at least and allows breakeven in a 224 kV, 10 MJ PF working with deuterium-tritium. We can suppress filamentation by superimposing a radial magnetic field to the interelectrode region of the PF where rundown occurs. A conservative estimate shows that a radial magnetic field $ \approx $ 1.4 T is enogh to suppress many known filamentation-triggering instabilities. Suitably located magnets (permanent, high-temperature-superconductors ('HTS')...) can generate this field; their lay-out resembles the lay-out of the sources of radial magnetic field in the cylindrical geometry of a Hall thruster for space propulsion; in HTS, the required current density is too small to trigger quenching.}


%%\pacs[JEL Classification]{D8, H51}

%%\pacs[MSC Classification]{35A01, 65L10, 65L12, 65L20, 65L70}

\maketitle

\section{The problem}\label{Introduction}

\begin{figure} 
\includegraphics[scale=0.3]{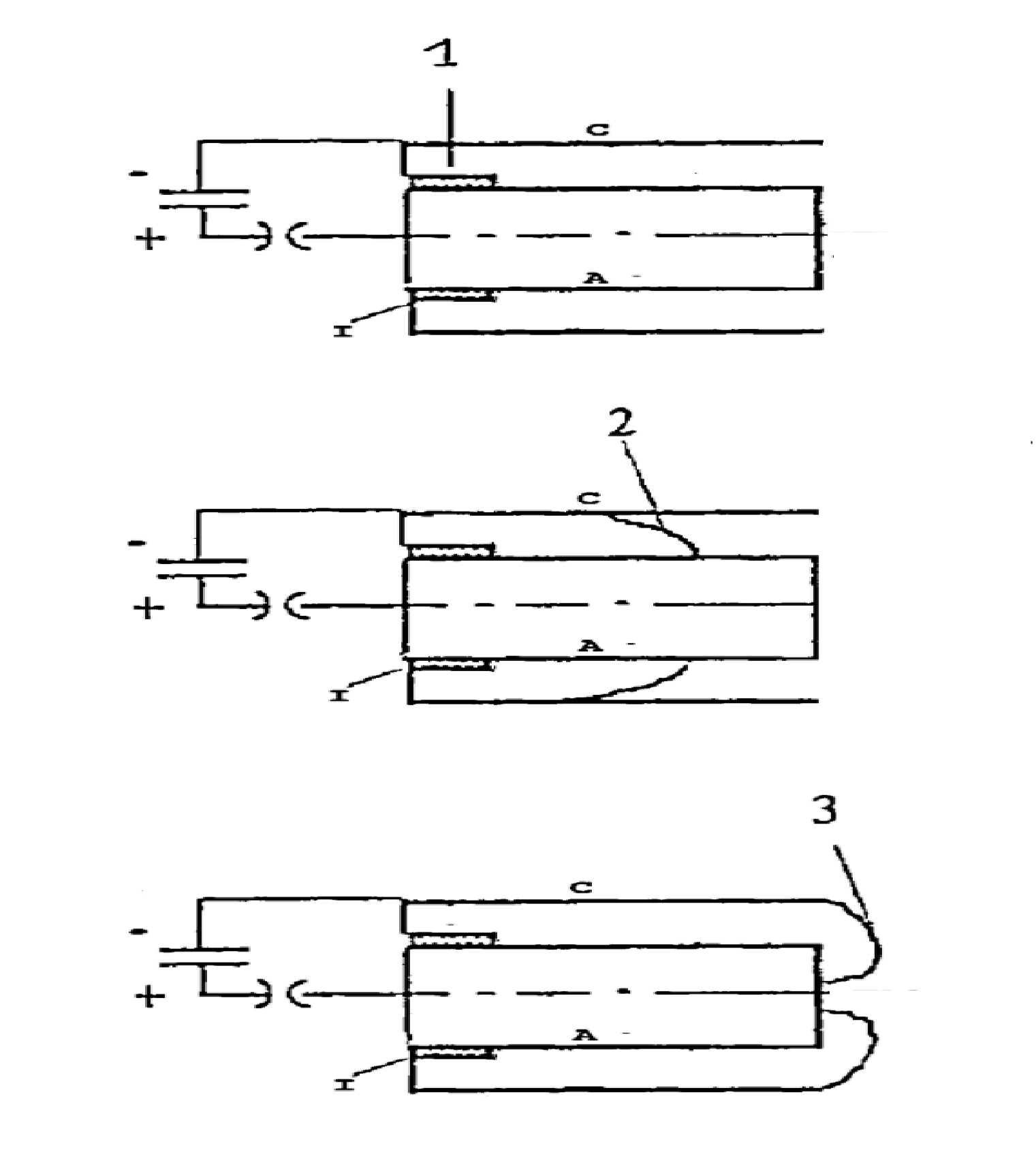}
\caption{\textit{Lay-out of a Mather-type PF (cross section). A: anode. C: cathode. I: insulator. Top: the sheath forms in 1, near the insulator. Centre: rundown. The sheath 2 moves towards the open end of the anode. Bottom: the sheath 3 collapses onto the symmetry axis and gives birth to a pinch (not displayed here).}}
\label{PF} 
\end{figure}

The '(Dense) Plasma Focus' ('PF') has been invented independently in USSR and in the US in the framework of the research on controlled nuclear fusion. PFs have been investigated for decades and still puzzle researchers \cite{MATHER0} \cite{MATHER} \cite{SCHOLZ} \cite{AULUCKSADOWSKIMIKLASZEWSKI}. In a popular ('Mather') configuration (Fig.~\ref{PF}), a typical PF device consists of two cylindrical electrodes, an inner anode - usually made of copper or other conductive materials - and an outer cathode, which may be constructed as a set of concentric cylindrical rods or as a tubular structure surrounding the anode. The gap between these electrodes forms the discharge area. An insulator sleeve, typically made from high-dielectric-strength materials such as ceramics (e.g., alumina) or pyrex, is positioned between the electrodes to prevent arcing until the system is operational. The insulator connects the electrodes mechanically at one end; electrodes are left open at the other end. The entire system is housed in a vacuum chamber filled with a low-pressure working gas, such as deuterium, tritium, or a mixture of both. The working gas is supplied to the vacuum chamber, and its pressure is carefully regulated to ensure optimal operation.\\

The device is powered by a high-voltage capacitor bank, which stores and releases a large amount $E $ of energy in a short burst to create the plasma. A spark gap or a similar triggering mechanism initiates the rapid discharge of the capacitor bank through the electrodes. The operational cycle begins when the capacitor bank discharges rapidly through the electrodes, creating a high-current, low-voltage arc in the working gas. This arc ionizes the gas, forming a plasma which can then conduct electricity. A plasma sheath is initially formed as a surface discharge over the insulator. The initial sheath gets detached from the insulator and gets accelerated by the magnetic force due to the interaction between the electric current and the magnetic field, and travels therefore along the electrodes towards the open end ('rundown'). The rundown of the sheath is the propagation of a shock which collects and ionizes the gas molecules ahead of the front. On its arrival at the open end, the plasma starts to accelerate towards the axis, driven by the magnetic force. Eventually, the sheath clashes on the axis in the form of a small, dense, hot, roughly cylindrical plasma - the 'focus', similar to a Z-pinch \cite{AreviewofthedenseZpinchHAINES} - in front of the central anode. After the pinch collapses, the system returns to its initial state, allowing the cycle to be repeated.\\

Today, $ E $ is in the range $ 10^{-1} $ J \cite{PAVEZ} $ \div 1.064 \cdot 10^6$ J \cite{SCHOLZ}. The corresponding ranges of the electric current $ I_{pinch} $ flowing across the pinch and of the anode radius $ a $ are 5 kA $ \div $ 2.3 MA and 0.8 mm $ \div $ 113 mm respectively. On the upper limit of these ranges (1.064 MJ, 2.3 MA, 113 mm) the minimum pinch radius is of the order of 1.5 mm. Estimates in smaller PFs do not differ so much. Typically, pinch density and temperature are $ \approx 10^{25} $ m$ ^{-3} $ and $ \approx $ keV respectively. Characteristic pinch time-scales take values from several to hundreds of nanoseconds.\\
 
Not surprisingly, instabilities grow and energy gets dissipated. In the pinch, these processes are correlated with the emergence of accelerated electrons and ions, hard X-ray pulses, as well as fusion-related neutron emission pulses if the working gas contains deuterium. These pulses are related either to beam-plasma fusion \cite{BRZOSKO} or to thermonuclear fusion \cite{RAGER} \cite{PPCFSOTO} reactions. The transfer of energy from the magnetized plasma to ions and electrons is a process not fully understood even today.  \\

When PFs are tuned in order to maximize the total emission $ Y_n $ of neutrons in a 100 \% deuterium plasma, their stunning feature is the validity of scaling laws which link macroscopic properties (anode radius, energy of the capacitor bank, rate of nuclear reactions) across seven orders of magnitude of $E $. According to Ref. \cite{PPCFSOTO}, for example, both relationships:\\
\begin{equation}
\label{scalingS}
68 \leq S \equiv \frac{I_{peak} (kA)}{a (cm) \sqrt{p_0 (mbar)}} \leq 95 
\end{equation}
\begin{equation}
\label{scalingepsilon}
10^{10} \leq \varepsilon \equiv \dfrac{28E (J)}{a(m)^3}  \leq 10^{11}
\end{equation}\\
($ S $ 'drive parameter', $ \varepsilon $ 'density energy parameter') apply to ten different PFs with $ 50 \mbox{ J} \leq E \leq 1.064 \mbox{ MJ}  $; here $ I_{peak} $ and $ p_0 $ are the peak current achieved during the discharge and the gas filling pressure respectively. Moreover, many experiments on PFs with $ E \ll $ 0.5 MJ lead to \cite{SCHOLZ}:\\
\begin{equation}
\label{scaling0}
Y_n \approx 10^7 \cdot E(kJ)^2
\end{equation}
and to \cite{LEE} \cite{DAMIDEH}:\\
\begin{equation}
\label{scaling}
Y_n = 1.8 \cdot 10^{10} I_{peak}(MA)^{3.8}
\end{equation}\\
The experimental observation that $ I_{peak} $ itself increases with increasing $ E $ allows \eqref{scaling0} to agree with \eqref{scaling}. As for \eqref{scaling}, it agrees - qualitatively at least - with physical intuition as $ I_{pinch} \approx \frac{2 I_{peak}}{3} $ \cite{NUKULIN}: the larger $ I_{peak} $, the larger $ I_{pinch} $, the larger the magnetic field in the pinch, the better the magnetic confinement of the plasma, the larger the number of  nuclear reactions in the pinch. Admittedly, various exponents are found in the literature for the power on the R.H.S. of \eqref{scaling}; since all of them are near 3.8, however, we stick to the latter value to keep ourselves consistent with the discussion of Ref. \cite{LEE} - see below.\\

Physically, \eqref{scalingS}, \eqref{scalingepsilon}, \eqref{scaling0} and \eqref{scaling} suggest that - as complex as it is - the physics is basically the same for most PFs investigated so far. For example, the volume of the vacuum chamber is basically proportional to $E $; its value is 20 dm$ ^{3} $ at 7 kJ \cite{ANGELI}. The corresponding maximum value of the plasma current is 0.39 MA - see Tab. 3 of Ref. \cite{PPCFSOTO}. Thus, a PF allows us to create a Z-pinch \cite{AreviewofthedenseZpinchHAINES} with current of a fraction of MA with the help of a cheap, table-top device. Even the smallest PF has essentially the same dynamic characteristics as larger machines; a small table-top-sized PF produces essentially the same pinch characteristics (temperature and density) as the largest PF. Thus, fusion reactions are even possible to be obtained in ultraminiature, $ E = 10^{-1} $ J devices, as they are in the bigger, $ E = 10^{7} $ J devices. However, the larger the PF, the larger the pinch volume with a corresponding longer lifetime, the larger both neutron and radiation yields.\\

Now, should \eqref{scaling0} hold also above 0.5 MJ and should we replace a pure deuterium gas with a 50\% - 50\% deuterium-tritium ('DT') mixture (where fusion reactions are $ \approx 80 $ times more probable and deliver 14.1 MeV neutrons), then we would achieve 'breakeven' at $ E \approx 400$ kJ; breakeven is the condition where the ratio $ Q $ ('fusion gain') between the energy obtained by fusion and the energy supplied by the external world to the plasma is = 1. (At a first glance, this conclusion seems to be a violation of Lawson criterion \cite{LAWSON}, since the spatially averaged temperature of the pinch is too low for DT breakeven in all PFs, i.e. $ \approx $ 1 keV rather than $ \approx $ 10 keV as required by Lawson criterion for DT fusion. But then, plasmoids - coherent structures of plasma and magnetic field - can form spontaneously in the pinch region and reach 7.5 keV \cite{AULUCKSADOWSKIMIKLASZEWSKI} \cite{SADOWSKI}. Moreover, Lawson criterion does not directly take into account beam-plasma fusion). The PF attracted interest from large national laboratories in the early days of controlled fusion research mainly because of \eqref{scaling0} \cite{RAGER}. There was hope that these experimental results could be scaled up to a very intense neutron source producing neutron fluxes similar to a fusion reactor \cite{ZUCKER} and even up to breakeven \cite{BRZOSKO}. As for fusion energy production for the grid, a necessary condition is $ Q > 1 $.\\

Unfortunately, however, experiments showed that if $E \gtrapprox $ 0.5 MJ then $ Y_n $ grows no more ('saturation') \cite{HEROLD} \cite{NUKULIN}. At $E =$ 0.5 MJ $ Y_n $ never exceeded $ 8 \cdot 10^{11} $ in pure deuterium \cite{SCHOLZ}. In DT this would give 147 J in output, i.e. $ Q \equiv Q_{record} = 3 \cdot 10^{-4} \ll 1$ (Admittedly, this is just a rule-of-thumb estimate. Detailed comparison between fusion rates in pure deuterium and DT requires comparison between the energies of the involved particles in the pinch and the cross-sections of the relevant fusion reactions \cite{DAMIDEH}. However, our estimate is enough for the purpose of our discussion). No detailed, satisfactorily description of the microscopic processes underlying saturation is yet available - see Ref. \cite{AULUCKSATURATION} for a review. Eventually, many large PF laboratories encountered failure of neutron yield scaling and discontinued their research. The evolution of PF as a research discipline was profoundly affected by this realization \cite{AULUCKSADOWSKIMIKLASZEWSKI}. Moreover, the failure of neutron yield scaling affected adversely both attempts to take advantage of the energetic particle beams and of the radiation produced in the pinch for non-fusion applications \cite{KRISHNAN}, and optimistic conceptual engineering studies of PF-based space propulsion for interplanetary and deep space missions \cite{HARDY} \cite{THOMAS}.\\ 

During rundown, limited efficiency of sweeping of the working gas with the sheath can allow part of the working gas to remain in the vicinity of the insulator; as a consequence, shunting of the total current in the PF system out of the pinch occurs, thus reducing the current actually flowing in the pinch and spoiling plasma confinement. On a qualitative basis at least, this shunting is today's commonly accepted explanation of saturation \cite{NUKULIN}. The problem is particularly severe precisely at large values of $ E $; we discuss it further in Sec. \ref{From suppression of filaments to breakeven}.\\

A non-exhaustive list of approaches followed to cope with saturation includes \cite{SCHOLZ}:
\begin{itemize}
\item Proper selection of the material the insulator sleeve is made of. As for the material, considerable increase in neutron yield has been e.g. achieved in the PF POSEIDON by replacing the Pyrex insulator by a ceramic tube \cite{AULUCKSADOWSKIMIKLASZEWSKI}. Experiments show that local variations in the insulator surface decrease the spatial uniformity of the sheath, leading to an azimuthally asymmetric focus and ,
reduced electron density; smoothly finished insulator surfaces are therefore preferred \cite{HOUSLEY}.
\item Proper selection of the geometry of the insulator sleeve. To start with, a proper choice of the insulator sleeve length is helpful \cite{HAHN}. Moreover, experiments show that addition of a field distortion element ('FDE') - namely, a metallic knife edge which encircles the base of the insulator sleeve - multiplies $ Y_n $ by 3 and by 5 according to Ref. \cite{NARDI} and Ref. \cite{NARDI2} respectively. Even more important: reproducibility of PF measurements may be poor \cite{KRISHNAN}, but FDEs dramatically reduce the fluctuation of $ Y_n $ from one shot to another \cite{NARDI2}. The model of Ref. \cite{AULUCKSATURATION} - a sophisticated refinement of the model of Ref. \cite{GRATTONVARGAS} - provides us with a further example, where a detailed description of the curved geometry of the sheath is available and where it turns out that a 60\% reduction of the radial size of the insulator can multiply $ Y_{n} $ by 200. However, the model relies on the rather overoptimistic assumption that (in the author's own words) \textit{the capacitor bank is fully discharged at the time when the plasma arrives at the anode centre}.
\item Modifications of the anode shape. Higher yields have been achieved after various iterations of the anode shape suggested by MHD-kinetic modeling techniques of the pinch implosion \cite{GOYONMJOLNIR}. A composite anode was used (a one piece anode with the diameter reduced at the end) was used and an enhancement of the order of 12?16\% in $ Y_n $ was reported \cite{PPCFSOTO}.
\item Monolithic (solid, single-piece) electrodes. They help eliminate arcing between separate pieces of the electrodes. This arcing is a significant source of impurities that leads to sheath disruption  \cite{LERNERCONFINEMENT}.
\item Corona-discharge pre-ionization at the start of the pulse. It generates a higher initial density of free electrons. This reduces the maximum electric fields and prevents the acceleration of runaway electrons, which otherwise cause electrode material vaporization \cite{LERNERRUNAWAY}. Alternatively, a radioactive source of electrons ($ ^{63} $Ni) placed near the insulator also provides pre-ionization.
\item Using tungsten for the electrodes, particularly the anode, reduces erosion and vaporization compared to copper. Tungsten is more resistant to the thermal shocks caused by runaway electron bombardment \cite{LERNERCONFINEMENT}.
\item Reduced anode radius. Given the values of $ S $ and $ I_{peak} $, the smaller $ a $ the larger $ p_0 $, the more unlikely the creation of runaway electrons \cite{LERNERRUNAWAY}.
%\item Decades of research have shown that the physics of the pinch - where neutrons are produced - is more complicated than previously envisaged. This may jeopardize accurate predictions of $ Y_{n} $. Reproducibility of measurements is often poor \cite{KRISHNAN}.
\end{itemize}
The US private company LPPFusion - the only one organization which explicitly aims at a commercially relevant PF fusion reactor to date - puts forward a different approach \cite{LERNER}. LPPFusion assumes that nuclear reactions occur in dense plasmoids where the magnetic field is so large \cite{BOSTICK} that quantum mechanical effects dramatically reduce radiation losses by hindering the collisional exchange of energy between ions (which undergo fusion) and electrons (which are the main responsible e.g. for Bremsstrahlung losses), so that even aneutronic fusion reactions p-$^{11}$B become possible. LPPFusion wants to optimize the production of plasmoids with the desired features. After suitable modification of the electrode design, preionization of the working gas before the main discharge, utilization of beryllium electrodes and addition of a modest axial magnetic field ($ \approx 10^{-4} $ T) with an external helical coil in order to drive a rotation of the sheath, LPPFusion claims to have obtained on May the 23$ ^{rd} $, 2016 $ Y_n = 2.5 \cdot 10^{11} $ in pure deuterium - corresponding to 0.2 J energy release - with $E = 60 $ kJ, a result seven times larger than predicted by the familiar scaling discussed above. LPPFusion claims also to have measured extreme ion temperatures - i.e. above 200 keV \cite{LERNERCONFINEMENT} - just like in the wire-array Z-pinch \cite{AreviewofthedenseZpinchHAINES} experiment of Ref. \cite{HAINES2006}. However, the tenet of LPPFusion - the beneficial occurrence of quantum effects in the plasmoids - has been questioned \cite{AULUCKSADOWSKIMIKLASZEWSKI} as the relevance of these quantum mechanical effects to a PF plasma is uncertain \cite{BAALRUD}. Moreover, breakeven in PF p-$^{11}$B fusion requires both ion temperature $ \gg $ electron temperature and the reflection and absorption of at least 50\%$ \div $60\% the power lost by radiation \cite{THOMAS} \cite{DIVITAPB}.\\

Whether for lack of funding of for lack of scientific knowledge, however, saturation still stands as an unresolved problem in spite of all efforts. When it comes to investigating saturation, the intricacies of the problems listed above and the huge variety of the spatiotemporal scales involved suggest we look for useful information with the help of some simple but reliable model. The widely accepted, macroscopic, simple but robust Lee's model \cite{LEE2014} - the workhorse of PF research - is right for us. Lee's model describes the rundown phase as the discharge of a capacitor bank with capacity $ C $, voltage $ V $ and energy $ E = \frac{CV^2}{2} $ on an underdamped RLC series circuit \cite{N214}. Lee's model \cite{LEE2014} allows us to compute $ I_{peak} $ - hence $ Y_n $, through \eqref{scaling} - once the inductive and the resistive impedances are known. In Lee's model, saturation is due to the increasingly dominant contribution of the time derivative of the plasma inductance in the rundown phase to the overall PF impedance as $ C \propto E $ increases \cite{LEE}. (The impact of inductance on saturation has been independently investigated also in Ref. \cite{NUKULIN}, leading to results somehow related to those of Ref. \cite{LEE}). The time derivative is an input constant in this description, which we outline in Sec. \ref{The problem}. Starting from a simplified, analytical description \cite{DIVITA1993} of the filamentary structure of the sheath in the rundown phase observed in experiments \cite{BOSTICK} \cite{AULUCKFILAMENTATION}, we compute this time derivative in Sec. \ref{A model for saturation}. Then, we show in Sec. \ref{From suppression of filaments to breakeven} how suppression of filamentation allows achievement of breakeven in a PF. We discuss briefly a method to achieve this suppression in Sec. \ref{How to suppress filaments}. Conclusions are drawn in Sec. \ref{Conclusions}. We focus on a 100\% deuterium plasma in the following, unless stated otherwise.
\section{The rundown} \label{The problem}
We focus our attention on the rundown in the following. During rundown, the plasma moves from the insulator end towards the open end of the anode, sweeping up neutral gas and forming a hot, dense layer that eventually creates the pinch. We consider the simplest representation in which the sheath is shown to go from the anode to the cathode perpendicularly. Experiments \cite{KWEK} show that there is actually a canting of the sheath and also that only a fraction typically 0.7 of the total current participates in driving the sheath. These points are explicitly accounted for in the model or Ref. \cite{LEE2014}; following Ref. \cite{LEE}, however, we do not consider these two effects below for simplicity. We show in Sec. \ref{From suppression of filaments to breakeven} that this simplifying assumption is too pessimistic.\\

We write the PF inductance $ L $ of a PF with outer electrode radius $ b $, anode length $ h $ and a thin sheath at distance $ z= z(t), 0 \leq z \leq h $ from the insulator located at $ z = 0 $ as $ L = L_{ext} + L_t $ where the constant quantity $ L_{ext} $ includes the inductances of the capacitor, transmission line, connections and switches, $L_t = \frac{\mu_{0}}{2 \pi} \ln (\frac{b}{a}) z (t) $ is the growing inductance of the discharge chamber and $ \mu_{0} = 4 \pi \cdot 10^{-7}$ $ T \cdot A^{-1} \cdot m $. Here the word 'thin' stands for 'with thickness $ \ll h$'. Since we pinch on the rundown phase before the sharp spike in the voltage signal which is usually taken as indication of the proper operation of the device, we invoke no sudden change of inductance in our discussion. However, $ \frac{dz}{dt} \neq 0 $ implies $ \frac{dL}{dt} \neq 0 $. Remarkably, a discussion of $ \frac{dL}{dt} $ during rundown is free from the ambiguities which a similar discussion concerning the pinch stumbles upon due to the complicated and time-varying geometry of the pinch \cite{AULUCKREPRESENTATION}. It is useful to define the 'dynamic resistance' $ D_R \equiv \frac{1}{2}\frac{dL}{dt}$ and obtain:
\begin{equation}
\label{DR}
D_R = \dfrac{\mu_{0}}{4 \pi} \ln \left( \dfrac{b}{a} \right) \cdot v \quad ; \quad v \equiv \dfrac{dz}{dt}
\end{equation}\\
According to Ref. \cite{LEE}, experiments show that on switching, as the capacitor discharges, the current rises towards its peak value, the current sheet is accelerated, quickly reaching nearly its peak speed, and continues accelerating slightly towards its peak speed at the end of the axial phase. Thus, $ \frac{dv}{dt} \approx 0 $ for most of the rundown. Correspondingly, we are going to take $ D_R = $ const. below.\\

The energy balance of the sheath reads:\\
\begin{equation}
\label{energybalance}
I V_{PF} = \dfrac{d}{dt}\left( \frac{L_t I^2}{2} \right) + P \quad ; \quad V_{PF} \equiv \dfrac{d (L_t I)}{dt}
\end{equation}\\
where we neglected the capacitance of the sheath \cite{N214}. The L.H.S. is the Poynting flux of electromagnetic energy delivered to the sheath; $ I $ is the current flowing across the plasma sheath and $ V_{PF} \neq V $ because of the impedance of the electrical circuit - with inductance $ L_{ext} $ and resistance $ R_{ext} $ - linking the discharge chamber with the capacitor bank \cite{N214}. On the R.H.S., $ P $ is the sum of the time derivative of the kinetic energy and the dissipated power. As $ \frac{dv}{dt} \approx 0 $, the former term is constant and strictly positive, as the advancing sheath sweeps the gas in front of itself. The latter term includes the contributions $ P_{\Omega} $ and $ P_V $ of Ohmic and viscous heating respectively as well as the net amount of energy exchanged per unit time due to ionization, recombination etc. For mathematical simplicity we neglect radiation; we are going to drop this unphysical assumption in Sec. \ref{A model for saturation}. After straightforward algebra, \eqref{energybalance} leads to:\\
\begin{equation}
\label{DRP}
D_R I^2 = P
\end{equation}\\ 
which shows that $ D_R $ plays the role of effective resistance of the sheath. The total amount of dissipated power in the PF as a whole is therefore equal to $ \left( R_{ext} + D_R \right) I^2 $. Typically, $ D_R = 7 $ m$ \Omega > R_{ext}$ for all PFs \cite{LEE}; for example, $ R_{ext} = 2.6 $ m$ \Omega $ on PF1000 \cite{LEESAW}. Then, we neglect $ R_{ext} $ in comparison with $ D_R $ below. We discuss two cases in the following: $ L_{ext} \gg L_t $ and $ L_{ext} \ll L_t $; we shall see that both cases lead basically to the same result, equation \eqref{QM} below, which will be the starting point of further investigation. If $ L_{ext} \gg L_t $ then Kirchoff's circuit laws in the underdamped RLC circuit which describe the PF as a whole \cite{N214} lead to the following result about $ I_{peak} $:\\
\begin{equation}
\label{Kirchoff}
I_{peak} = V \sqrt{\dfrac{C}{L_{ext}}} \exp \left[ - \dfrac{\alpha}{\omega_d} \arctan \left( \dfrac{\omega_d}{\alpha} \right)\right] \quad ; \quad \alpha \equiv \dfrac{D_R}{2L_{ext}} \quad ; \quad \omega_d \equiv \sqrt{\dfrac{1}{LC} - \alpha^2}
\end{equation}\\
We assume $ D_R \ll \sqrt{\frac{L_{ext}}{C}}$ in agreement with Ref. \cite{LEE}, so that \eqref{Kirchoff} reduces to:\\
\begin{equation}
\label{Ipeak}
I_{peak} = \dfrac{V}{\sqrt{\frac{L_{ext}}{C}}+\frac{\pi}{4}D_R} = \dfrac{4V}{\pi D_R}\dfrac{1}{x^{-\frac{1}{2}}+1} 
\end{equation}\\
where $ x \equiv \frac{E}{E_c} $ and $ E_c \equiv 8L_{ext} \left( \frac{V}{\pi D_R}\right)^2 $. Equations \eqref{scaling} and \eqref{Ipeak} give:\\
\begin{equation}
\label{yield}
Y_n \propto \left( \frac{E_c}{L_{ext}} \right)^{1.9} (x^{-\frac{1}{2}}+1)^{-3.8}
\end{equation}\\
as well as the following relationship for the fusion gain $ Q \propto \frac{Y_n}{E} $:\\
\begin{equation}
\label{Q}
Q \propto \frac{E_c^{0.9}}{L_{ext}^{1.9}} f(x) \quad ; \quad f(x) \equiv x^{-1} \cdot (x^{-\frac{1}{2}}+1)^{-3.8}
\end{equation}
(If $ x \ll 1 $ then \eqref{yield} gives $ Y_n \propto E^{1.9} $, which is slightly more pessimistic than \eqref{scaling0}. Conservatively, we stick to \eqref{yield} below). According to \eqref{Q}, $ Q $ attains its maximum value $ Q_{\max} $ at $ x \equiv x_0 = 0.81 $ , where:\\
\begin{equation}
\label{QM}
Q_{\max} \propto \dfrac{E_c^{0.9}}{L_{ext}^{1.9}} \propto \dfrac{V^{1.8}}{L_{ext} D_R^{1.8}}
\end{equation}\\ 
If $ x \ll x_0$ then $ Q \propto \frac{E^{0.9}}{L_{ext}^{1.9}} $ and we can easily raise $ Q $ by raising $ E $. If $ x = x_0$ then $ Q $ stops increasing ('saturation'). If $ x \gg x_0 $ then $ Q $ decreases with increasing $ x $.
Physically, a capacitor bank of voltage $ V $ discharging into a constant resistance such as $  D_R $ will have a peak current $ I_{peak} $ approaching an asymptotic value of $ I_{peak} \propto \frac{V}{D_R}$ when the bank capacitance $ C $ is increased to such large values that $ \sqrt{\frac{L_{ext}}{C}} \ll  D_R$. Thus, $ D_R $ causes saturation of $ I_{peak} $ \cite{LEE}, hence of $ Y_n$ because of \eqref{scaling}. (Rigorously speaking, this explanation is qualitative only, as \eqref{Ipeak} holds in the $ \sqrt{\frac{L_{ext}}{C}} \gg D_R $ limit only; however, it gets to the heart of the topic). We have seen in Sec. \ref{Introduction} that the maximum value of $ Y_n $ obtained so far in pure deuterium would correspond to $ Q_{\max} = Q_{record} \ll 1$ in a DT plasma. Thus, no PF achieved breakeven so far.\\ 

According to Ref. \cite{LEE}, most PF experiments to date operate in a narrow range of voltages of $ 15 \div 50 $ kV. Only the PF SPEED-II operated at hundreds of kilovolts. But SPEED-II has a large $ \sqrt{\frac{L_{ext}}{C}} $, which itself limits the current. As a matter of principle, we may still operate a range of PFs at fixed $ V $ with ever larger $ E $ until the impedance becomes negligible due to the very large value of $ C $. Admittedly, the saturation effect is still there, but at saturation $ I_{peak} \propto \frac{V}{D_R} $. Thus, by raising $ V $ we can raise $ Q_{\max} $ and overcome today's saturation. In a nutshell, this is the proposal outlined in the last lines of Ref. \cite{LEE}.\\

As for the case $ L_{ext} \ll L_t $, we note that - under optimum conditions - the rundown must end just at the instant of the maximum current. Now, an increase in $ E $ at a constant $ V $ voltage of the capacitor bank is usually achieved by increasing $ C $. But an increase in $ C $ increases the typical discharge time-scale of the condenser bank $ \tau_{LC} \left( \propto \sqrt{C} \right) $ and inevitably causes one to lengthen $ h $ in order to provide the synchronism required above. If $ E $ is large enough, therefore, $ L_t \propto h \ln \left( \frac{b}{a} \right) > L_{ext} $. In the authors' own words, according to Ref. \cite{NUKULIN} \textit{the current amplitude is determined by the chamber inductance rather than by the inductance of the capacitor bank. A further increase in the capacitance of the energy storage bank does not lead to an increase in the discharge current because of the growth of the chamber inductance. Therefore, the current and, accordingly, the neutron yield saturate.} Not surprisingly, \textit{a further increase in the neutron yield can be achieved by increasing the charging voltage of the capacitor bank} - much like Ref. \cite{LEE}, and for similar reasons.\\

Alternatively, we may lower $ L_t $ by reducing $ \frac{b}{a} $. We do not delve further into the issue here; in the following, we limit ourselves to investigate the dependence of $ Q_{\max} $ on $ V $ with the help of \eqref{QM} and, to this purpose, we assume the PF geometry to be already optimized. This assumption is reasonable as the fundamental result of Ref. \cite{NUKULIN} - its equation (7) - gives $ I_{peak} \propto \frac{V}{v \ln \left( \frac{b}{a} \right)} $, which in turn is $\propto \frac{V}{D_R}$ - because of \eqref{DR} - and is in agreement with \eqref{Ipeak}. (Correspondingly, should we follow equation (10) of Ref. \cite{NUKULIN} we should replace our $ f(x) $ in \eqref{Q} with another unimodal function of $ x $ with another value of $ x_0 $). This means that the conclusions of Ref. \cite{NUKULIN} agree with Ref. \cite{LEE} for the purposes of our discussion: equation \eqref{QM} holds in both discussions of Ref. \cite{LEE} and Ref. \cite{NUKULIN}. Even if the model of Ref. \cite{NUKULIN} has been heavily criticized \cite{LEE2008}, therefore, we may safely rely on \eqref{QM} in the following.\\

Indeed, the validity of \eqref{QM} regardless of the actual value of $ \frac{L_t}{L_{ext}} $ has a simple explanation. Consider what happens as $ L_{ext} $ is varied for a fixed $ \frac{dL_t}{dt} $ impedance. The moving current sheet has inertia, i.e., it lifts off the insulator slowly and accelerates only after a certain delay after breakdown. If $ L_{ext} $ is too low, the current rapidly rises before the sheet begins to move, the sheet is accelerated by a high value of current, and the high impedance $ \frac{dL_t}{dt} $ damps the current. The peak is reached before the sheet reaches the end of the anode, and the pinch phase is driven by a lower current than the peak current $ I_{peak} $. On the other hand, if $ L_{ext} $ is too large then the rate of rise of current is slow, and that allows the sheet to accelerate while keeping the current high (peak reached at the end of the anode), but the magnitude of that peak current is lower due to the higher value of $ L_{ext} $. There is obviously a value of $ L_{ext} $  that keeps the current from dipping before the sheet has reached the end of the anode, while maximizing its value. Under optimum condition, we suppose we have found this optimum value of $ L_{ext} $, and keep it fixed in the following \cite{KRISHNAN}.\\

Let us check if the proposals of Refs. \cite{LEE} and \cite{NUKULIN} of raising $ V $ are practical. To this purpose, let us compare the values $ Q_1 $ and $ Q_2 $ of $ Q_{\max} $ in two PFs with voltages $ V_1 $ and  $ V_2 $ respectively, with the same values of $ L_{ext} $, as discussed above. Following Ref \cite{LEE}, we assume also $ D_R $ to be the same. Then, \eqref{QM} gives:\\
\begin{equation}
\label{comparisonwithfilament}
\dfrac{Q_2}{Q_1} = \left( \dfrac{V_2}{V_1} \right)^{1.8}
\end{equation}\\
To fix ideas: if a PF with $ V_1 = $ 50 kV achieves $ Q_1 = Q_{record} = 3 \cdot 10^{-4}$ then we need $ V_2 = $ 4.5 MV in order to achieve breakeven ($ Q_{2} = 1 $). If $ Q_{record} $ is obtained at $ E = $ 0.5 MJ then breakeven corresponds to $ E = $ 4.05 GJ - a value which is definitely far from practically achievable.\\

Actually, \eqref{QM} allows us to obtain the same increase in $ Q_{\max} $ with smaller voltage increase if we reduce $ L_{ext} $. However, this is easier said than done, because $ L_{ext} $ is already minimized as much as possible already in the design phase of a PF under optimum conditions. A question naturally arises: what about $ D_R $? To get an answer, we need more information about $ D_R $. In other words, we need a a model for saturation. We have dealt with three quantities ($ P $, $ D_R $ and $ I_{peak} $) with the help of two balance equations: the balance of energy \eqref{energybalance} (which provides us with the link \eqref{DRP} between $ P $ and $ D_R $) and the balance of electric charge (through Kirchhoff's circut equations, which in turn lead to \eqref{Ipeak}). In the discussion of Ref. \cite{LEE}, the value of $ D_{R} $ is an input quantity; a full description of saturation needs a third, independent balance equation.

\section{A model for saturation} \label{A model for saturation}
We have seen that $ D_R $ is related to the growth of $ L $ during rundown. 
%(The fact that a conductor fed with a  given electric current spontaneously raises its own inductance whenever possible is far from surprising - think e.g. of a current-carrying solenoid which pulls a piece of iron towards itself). 
We have neglected the inner structure of the sheath so far. Experiments \cite{BOSTICK} \cite{ZORONDO} show that the structure of the sheath during rundown is made of radial filaments. (Even in the pinch, actually, the magnetic field has a complex three-dimensional ('3D') structure \cite{AULUCKPOLOIDAL}; the pinch is formed by a multiplicity of filaments \cite{NARDI2}. In the following, however, we focus on the plasma sheath during the rundown). Observation of filaments in the PF PF1000 allows conceptualization of filamentation as a native feature of the traveling current distribution behind an ionizing strong shock wave \cite{AULUCKFILAMENTATION}. The amount $ W_{sh} \equiv \int_{sheath} \frac{\vert \textbf{B} \vert^2}{2 \mu_0} d^3x$ of magnetic energy stored in the structure of the sheath - usually neglected in conventional, two-dimensional ('2D') descriptions of the sheath \cite{GRATTONVARGAS} \cite{POTTER} - may be significantly larger than the amount stored in a perfectly smooth, 2D system - and all the more as filaments interact magnetically with each other, leading to a further contribution to $ L $ due to these mutual interactions. In particular, we are dealing with 'vortex-current' filaments, which represent well-localized concentrations of both electric current density and fluid vorticity \cite{SOTO} \cite{AULUCKDESCRIPTION}. These filaments may be exact solutions of the equations of motion in a low-pressure plasma (like e.g. the sheath just after its birth near the insulator sleeve) embedded in a strong magnetic field \cite{BERGMANS}. In agreement with the conservation of angular momentum during the spontaneous formation of filaments, the sheath consists of a number of pairs of vortex-current-filaments running in the direction of increasing $ z $ between the inner and the outer electrode; each pair consists of one filament with positive radial electric current density and positive vorticity and another filament with positive radial electric current density and negative vorticity \cite{BOSTICKORIGINAL}. A rigorous treatment shows that interaction is possible between filaments with opposite sign of the dot product of fluid velocity and Alfven velocity \cite{PETVIASHVILI}. The magnetic interaction between the two filaments inside each pair greatly increases $ W_{sh} $ \cite{DIVITA1993}. The observed, spontaneous filamentation affects therefore $ L_t $ and $ D_R $.\\

Let us assume that the filaments in the sheath are 'quasi force-free' - i.e., approximately similar to Beltrami-like structures, with the electric current and the magnetic field almost parallel to each other, so that the related Lorenz force $ \approx 0 $. This assumption is far from new - it dates back to the work of W. Bostick \cite{BOSTICKORIGINAL} - and agrees with the observations reviewed in Ref. \cite{WHBOSTICK}. (Even in the pinch, the existence of quasi force-free structures has been hypothesized long ago \cite{BOSTICKORIGINAL} \cite{WHBOSTICK} \cite{SESTERO}, postulated in Ref. \cite{LERNER} and investigated - among many others - in Ref. \cite{DIVITAHOTSPOT}). If the assumption of quasi force-free filaments is correct, the amount of $ W_{sh} $ stored in the filamentary structure during rundown at the expense of the energy supply of the condenser bank is wasted from the point of view of the mechanical work done after the rundown upon the plasma in front of the anode when the pinch is formed. In this respect, we recall that experiments on the PFs POSEIDON and PF1000 show that the radial velocity of the plasma during the formation of the pinch is lower for discharges with lower $ Y_{n} $ \cite{AULUCKSADOWSKIMIKLASZEWSKI}.\\

A natural explanation for saturation seems therefore to be possible: if $ I $ becomes large enough then some physical process triggers the development of filamentary structure in the sheath which spoils the transfer of energy from the condenser bank to the plasma, as the quasi force-free magnetic fields of the filaments tend to drain the available energy while exerting no force upon the plasma. (Even beyond our approximation of a flat plasma sheath, experiments show that the plasma region containing the filaments moves away from the anode surface and is confined in a region of the plasma sheath, such as a toroidal plasma belt \cite{CASANOVA}, without reaching the anode top \textit{nor participating in the radial compression phase} \cite{ZORONDO}). As for the nature of these physical processes, they are likely to be electromagnetic instabilities, rather than purely electrostatic. This is far from surprising, as quasi force-free plasmas may have high values of the dimensionless ratio $ \beta $ of kinetic and magnetic pressure. The larger $ I $, the larger the magnetic energy available to these instabilities.\\

A physical quantity which is routinely utilized when it comes to investigating quasi force-free plasmas is the magnetic helicity $ K \equiv \int_{sheath} \textbf{A} \cdot \textbf{B} d^3x$ with $ \textbf{B} = \nabla \wedge \textbf{A} $ and $ \textbf{A} $ vector potential: its balance shall be the third, independent balance equation invoked in Sec. \ref{The problem}. The analytical discussion \cite{DIVITA1993} of the balance of $ K $ of a sheath made of quasi force-free filaments during rundown leads to the following results, which we are going to take advantage of below:
\begin{itemize}
\item $ W_{sh} $ increases with time during rundown; 
\item $ \left[ W_{sh} \right]_{I = I_{peak}} $ grows with increasing $ E $ at least as $ \propto E^2 $. It scales just as $ \propto E^2 $ in the extreme, most conservative case where the temperature of the sheath does not increase with increasing $ E $. This relationship follows from the requirement that the heating power supplied to the plasma must overcome the total radiative losses - neglected so far - in order to prevent radiative collapse. In deuterium (and DT) plasmas the only form of radiation we take into account is Bremsstrahlung.
\item saturation of $ Y_{n} $ occurs as the value $ \left[ W_{sh} \right]_{I = I_{peak}} $ of $ W_{sh} $ when $ I = I_{peak} $ is $ \approx $ the whole available amount $ \frac{LI_{peak}^2}{2} $ of magnetic energy. This means that all magnetic energy available to the sheath is eventually stored in the magnetic field of quasi-force free filaments and is therefore wasted. Since $ I_{peak}^2 $ is (roughly speaking!) $ \propto E $ and $ \left[ W_{sh} \right]_{I = I_{peak}} $ grows with increasing $ E $ at least as $ \propto E^2 $, it follows that $ \left[ W_{sh} \right]_{I = I_{peak}} \ll \frac{LI_{peak}^2}{2} \propto E$ for small $ E $ and becomes not negligible as $ E $ grows, thus leading to saturation at high $ E $. Alternatively, we may formally replace \eqref{scaling0} with $ Y_n \propto \left\lbrace E - \left[ W_{sh} \right]_{I = I_{peak}} \right\rbrace ^2 $, retrieve \eqref{scaling0} in the limit of small $ E $ and approach saturation as $ E $ starts growing. The results of Ref. \cite{DIVITA1993} agree with observations in the PF PF360.
\item The balance of $ K $ reads:\\
\begin{equation}
\label{helicitybalance}
\dfrac{d}{dt}\dfrac{W_{sh}}{\mu} + \dfrac{2 \eta_{\Omega} \mu W_{sh}}{\mu_{0}} - \dfrac{I V_{PF}}{2\mu} = 0
\end{equation}\\
The 1$ ^{st} $, the 2$ ^{nd} $ and the 3$ ^{rd} $ term on the L.H.S. stand for the time derivative of $ K $, the amount of $ K $ dissipated per unit time and the source term - i.e., the amount of $ K $ supplied per unit time by the external world - respectively. Here $ \eta_{\Omega} $ and $ \mu^{-1} $ are the resistivity and the minor radius of a single filament respectively. In the 2D case $ \mu^{-1} \rightarrow 0$, $ W_{sh} \rightarrow 0$ and $ K \rightarrow 0$.
\item We obtain a lower bound on the source term in \eqref{helicitybalance} under the conservative assumption $ I V_{PF} = P_{\Omega nf } $, to be discussed below. Here $ P_{\Omega nf } $ is the Ohmic power computed in a plasma sheath with no filamentation (quantities referring to such a sheath are labeled with the subscript 'nf' here and in the following). Under this assumption the relationship $ \mu^{-2} \equiv 2 \eta_{\Omega} \mu_{0}^{-1} W_{sh} P_{\Omega nf } ^{-1} $ holds.
\end{itemize}
In the following, we assume that the filamentary structure of the sheath is well-developed. Accordingly, we tale the limit $ \mu \rightarrow 0 $ (which is the opposite of the $ \mu^{-1} \rightarrow 0$ of the 2D case). In a scenario where filaments arise spontaneously during rundown, this assumption is reasonable near the end of this motion, i.e. when $ I \approx I_{peak} $. In this limit, \eqref{helicitybalance} leads to:\\
\begin{equation}
\label{Poh}
\dfrac{dW_{sh}}{dt} = \dfrac{P_{\Omega nf }}{3}
\end{equation}\\
Equation \eqref{Poh} has a far-from-trivial consequence precisely when $ I \approx I_{peak} $. Generally speaking, indeed, the definition $ D_R \equiv \frac{1}{2}\frac{dL}{dt}$ and the relationship $ W_{sh} = \frac{L I^2}{2}$ (which holds approximately at saturation when $ I \approx I_{peak} $) give $ \frac{dW_{sh}}{dt} = D_R I^2 + \frac{LI}{2}\frac{dI}{dt}$. When $ I \rightarrow I_{peak} $ then $ \frac{dI}{dt} \rightarrow 0 $ and $ \left[\frac{dW_{sh}}{dt}\right]_{I = I_{peak}} \rightarrow  D_R I_{peak}^2 = P_{I = I_{peak}}$ where we recalled that $D_R = $ const. and invoked \eqref{DRP}. But then \eqref{Poh} gives $ \left[\frac{dW_{sh}}{dt}\right]_{I = I_{peak}} = \left[ \frac{P_{\Omega nf }}{3} \right]_{I = I_{peak}} $, hence:\\
\begin{equation}
\label{Ppeak}
P_{I = I_{peak}} = \dfrac{1}{3}\left[ P_{\Omega nf } \right]_{I = I_{peak}}
\end{equation}\\
Now, the L.H.S. is the non-electromagnetic contribution to the energy balance when the sheath is endowed with a fully-developed filamentary structure. It includes Ohmic power $ P_{\Omega} + P_V $ among other contributions, hence $ \left[ P_{\Omega} + P_V\right]_{I = I_{peak}} \leq P_{I = I_{peak}} $ and \eqref{Ppeak} gives: $ \left[ P_{\Omega} + P_V\right]_{I = I_{peak}} \leq P_{I = I_{peak}} = \frac{1}{3}\left[ P_{\Omega nf } \right]_{I = I_{peak}} \leq \frac{1}{3}\left[ P_{\Omega nf } + P_{V nf}\right]_{I = I_{peak}}$, where $ P_{V nf} \geq 0 $ is the amount of power dissipated by viscosity when no filament is present. As a consequence, $ \left[ P_{\Omega} + P_V \right]_{I = I_{peak}} < \left[ P_{\Omega nf } + P_{V nf}\right]_{I = I_{peak}}$: filamentary sheaths dissipate \textit{less} power than sheaths where no filaments occur.\\

This conclusion fits nicely the fact that the relaxed states of a Hall magnetofluid \cite{TURNER} invoked for the description of the PF sheath \cite{AULUCKDESCRIPTION}:
\begin{itemize}
\item minimize $ P_{\Omega} + P_V $ \cite{DIVITAHOTSPOT}.
\item have a typical radial size $ \mu^{-1} \approx $ ion collisionless skin depth $d_i $ \cite{AULUCKFILAMENTATION}, in agreement with experiments which confirm a vortex-filament description \cite{SOTO} (in a MHD description, in contrast, filaments are infinitely thin \cite{PETVIASHVILI});
\item describe quasi force-free filaments \cite{DIVITAHOTSPOT} as the Hartmann number $ Ha $, namely the ratio of Lorentz force to the viscous force, is not too large. Equation (2.1) of Ref. \cite{DIVITAHARTMANN} gives $ 20 \leq Ha \leq 70$ for $ 0.1 \leq \beta \leq 1 $ and typical size $d_i $, as usual in quasi force-free structures; should no filamentation occur, the Lorenz force - hence $ Ha $ - would be much larger. Approximately at least, the electric current density in these filaments satisfies London equation, in agreement with Refs. \cite{TSYBENKONIKULIN} and \cite{TSYBENKOERISKIN}.
\end{itemize}
The latter result agrees with the outcome of Ref. \cite{LERNERCONFINEMENT}; in the authors' own words \textit{a heated plasma moves into a cool plasma or neutral gas in a way that minimizes the total dissipation of energy. The two principal ways how energy is dissipated from the sheath in a plasma pinch device are through electrical resistance of the current moving through the sheath, and through hydrodynamic friction of the sheath moving through the background medium.} In their discussion,  
%read $ P_V = \frac{\rho_{0} v^2}{2} \cdot \frac{vl_{f}}{\mu} \cdot N $ and $ P_V = \frac{\rho_{0} v^2}{2} \cdot \frac{vl_{f}}{\mu} \cdot N $  respectively in a system of $ N $ filaments - each of them with radius $ \frac{1}{\mu} $ and length $ l_{f} $ - which carries a total electric current $ I $, where $ \rho_{0} $, $ \frac{vl_{f}}{\mu} $, $ \eta_{\Omega} $ and $ \pi \mu^{-2} $ are the mass density of the filling neutral gas in front of the advancing sheath, the volume intercepted per unit time by one filament with cross section $ \frac{l_f}{\mu} $, the electrical resistivity and the cross-section of one filament respectively; moreover, the filaments are in parallel, i.e. subject to the same interelectrode voltage. As filamentation develops during rundown $ \mu $ increases; according to \eqref{PviscPOhm}, correspondingly, $ P_V $ decreases and $ P_{\Omega} $ increases. (Admittedly, we overlooked the contribution to $ P $ of Ohmic dissipation in the electrical circuits of the PF outside the sheath. However, we can reasonably assume that sheath filamentation affects this contribution only weakly). 
electrical dissipation is minimized by an even distribution of current, but hydrodynamic friction is minimized by filamentation.\\ 

The fact that filamentation raises Ohmic heating explains why our assumption $ I V_{PF} = P_{\Omega nf } $ above is a conservative one. The Poynting flux $ I V_{PF} $ is the source of the the filamentation-related growth of $ K $ in \eqref{helicitybalance} and supports not only Ohmic dissipation but also the growth of sheath inductance and the viscous dissipation in \eqref{energybalance}. By taking in account only Ohmic dissipation we obtain a lower bound on this source term, and an even lower bound is further obtained by replacing the Ohmic power with its own lower bound, i.e. its value $ P_{\Omega nf } $ in a filament-free sheath. Actually, our discussion underestimates filamentation. \\

The balance between electrical dissipation and hydrodynamic friction determines if filamentation will occur. The requirement that the sheath finds this minimum within the space available between the electrodes leads to a constraint on $ \eta_{\Omega} $; in turn, this constraint leads to the prediction that filaments in the sheath are only observed if the impurity content does not exceed a threshold - in agreement with observations \cite{LERNERCONFINEMENT}.\\ 

A further, independent consideration seems to strenghten our conclusion. Our discussion holds regardless of the fact that the inner and the outer electrode are the anode and the cathode respectively. Thus, its tenet - the occurrence of electromagnetic instabilities at high current which spoil the transformation of condenser bank energy into mechanical plasma energy through spontaneous formation of quasi force-free magnetic structures - is likely to hold also when the the inner and the outer electrode are the cathode and the anode respectively. Indeed, this seems precisely to be the case of magnetoplasmadynamic thrusters ('MPD') - see Ref. \cite{HARDY} for a comparison between PF and MPD. In MPD, the onset of thrust-spoiling instabilities occurs when collisions between ions and neutrals destabilize fast magnetosonic waves and triggered filamentation \cite{SPONTANEOUSSYMMETRYBREAKING}.\\

Finally, \eqref{DRP} and \eqref{Ppeak} allow us to write the value of $ D_R $ for our filament-affected rundown - formally at least: $ D_R = \frac{1}{3 I_{peak}^2}\left[ P_{\Omega nf } \right]_{I = I_{peak}} $.

\section{From suppression of filaments to breakeven}\label{From suppression of filaments to breakeven}
As shown in Sec. \ref{A model for saturation}, filamentation during rundown produces saturation; consequently, to prevent the latter we have to suppress the former. This conclusion may seem puzzling at first sight, as the 'squirrel cage' geometry made of many rods is the lay-out of choice for the cathode of PF at high $ E $ and seems uniquely fit to produce filaments in the sheath during rundown one filament per rod. However, we observe that the Lorenz force is maximum near the anode surface (this is actually a reason for the canting of the sheath we neglected), i.e. where the images of the sheath show that filaments are particularly well distinguishable \cite{BOSTICK}. To improve the impact of our efforts on the Lorenz force, therefore, it is near the anode surface where we want to suppress filamentation as much as possible, regardless of the detailed lay-out of the cathode.\\

Our approach is the opposite of LPPFusion's approach \cite{LERNER} \cite{LERNERCONFINEMENT} \cite{LERNERRUNAWAY}: they want to take advantage of filaments in order to create plasmoids in the pinch with large enough magnetic field while invoking quantum physics, we want to get rid of filaments in the rundown phase and rely on familiar scaling laws. Here we suppose we are able to get rid of those parasitic, quasi force-free fields somehow, and investigate the consequences of this suppression. To start with, we focus our attention on one PF. Then, we will compare the $ Q $'s obtained on two filament-free PFs at different $ V $'s, just like in Sec. \ref{A model for saturation}. We outline a possible approach to filament suppression in Sec. \ref{How to suppress filaments}.\\

Filaments reduce dissipation. Consequently, their suppression raises it; in particular, it raises both $ P_{I = I_{peak}} $ and $ v $ - because \eqref{DR} and \eqref{DRP} lead to $ P = \frac{\mu_{0}}{4 \pi} \ln \left( \frac{b}{a} \right) \cdot I^2 \cdot v \propto v $. This result suggests that suppression of filamentation may be beneficial, as experiments on the PF POSEIDON showed that the lower $ v $ the lower $ Y_n $ \cite{AULUCKSADOWSKIMIKLASZEWSKI}. At a first glance, this conclusion is puzzling; after all, we would expect dissipation to affect saturation negatively. On the other hand, however, filaments in the rundown phase spoil the transfer of energy from the condenser bank; their suppression increases therefore the amount of energy which is eventually available to the compression of the pinch in front of the anode. In particular, the kinetic energy of the sheath is $ \propto v^2 \propto P^2$: if we suppress filamentation then $ P $ increases, but this kinetic energy increases more.\\

The scaling $ P \propto v $ leads to a result which can be checked experimentally. Generally speaking, $ v \propto S $ \cite{PPCFSOTO} \cite{LEESERBAN}. It follows that $ S \propto P $, hence suppression of filamentation raises  $ S $. A quantitative estimate of the value $ S_{nf} $ of the drive parameter in a sheath with no filaments is  possible, as $ \frac{S_{nf}}{S} = \frac{\left[ P_{ nf } \right]_{I = I_{peak}}}{P_{I = I_{peak}}} $. Actually, \eqref{DRP} gives $ P_{I = I_{peak}} = D_R \cdot I_{peak}^2$
% with the value $ D_R = 7 $ m$ \Omega $ provided by Ref. \cite{LEE} in agreement with present observations
. Then, \eqref{Ppeak} gives $ \left[ P_{\Omega nf } \right]_{I = I_{peak}} = 3 \cdot D_R \cdot I_{peak}^2$. We obtain $ \left[ P_{ nf } \right]_{I = I_{peak}} $ by adding two positive quantities to $ \left[ P_{\Omega nf } \right]_{I = I_{peak}} $, namely the time derivative of the kinetic energy and the viscous power. 
%Conservatively, we neglect the time derivative of the kinetic energy altogether. This is in agreement with the estimate of Sec. 12A of Ref. \cite{DOLAN}, where the contribution of the kinetic energy to the energy balance of the sheath never exceeds $ 0.17 \cdot E $. As for the viscous power, it gets minimized in the well-developed filamentary structure at the end of the rundown, and we neglect it too. 
Then $ \left[ P_{nf } \right]_{I = I_{peak}} > \left[ P_{\Omega nf } \right]_{I = I_{peak}} $ and:\\
\begin{equation}
\label{S}
S_{nf} > 3S 
\end{equation}\\
This is remarkable, as Refs. \cite{PPCFSOTO} and Ref. \cite{LEESERBAN} propose the attractive scaling laws $ Y_n \propto S^4 \cdot I_{peak}^4 $ and $ Y_n \propto S^9 $ respectively; in both cases the thermonuclear component of $ Y_n $ outstrips the beam?target component.\\

Further discussion requires detailed knowledge of both the mechanical and the Ohmic contribution to $ P $ separately, with and without filaments. Such knowledge is not available; the reason is that the model of Ref. \cite{DIVITA1993} which our discussion stems from deals with the balance of $ K $ underlying \eqref{helicitybalance}, not with the balance of generalized helicity $ H = \int (\textbf{A} + d_i \textbf{v}) \cdot (\textbf{B}+ d_i \nabla \wedge \textbf{v}) d^3x$ ($ \textbf{v} $ fluid velocity) of a Hall magnetofluid \cite{TURNER} where the mechanical and the electromagnetic degrees of freedom are dealt with on an equal footing. In turn, this is because Ref. \cite{DIVITA1993} focusses on the smallness of Lorenz force in order to discuss saturation, and is therefore concerned mainly on the electromagnetic side of the problem. To the purposes of our discussion, however, suffice it to say that - at saturation - \eqref{Ipeak} and \eqref{DRP} show that $ I_{peak} \propto \frac{V}{D_R} $ and $ P_{I = I_{peak}} = \frac{V^2}{D_R} $ respectively.\\ 

Now, let us compare again the value $ Q_2 $ of $ Q_{\max} $ in a PF with voltage $ V_2 $ with the value $ Q_1 $ of $ Q_{\max} $ in a PF with voltage $ V_1 $, just like in Sec. \ref{The problem}. This time, however, filamentation is somehow suppressed in both cases 1 and 2. As a consequence, the universal filamentation mechanism which rules the dynamic resistance in all PFs \cite{LEE} is no longer at work, and we are no more allowed to assume that $ D_R$ is the same in 1 and 2. Still, filamentation leaves $ L_{ext} $ unaffected; then, its suppression too leaves $ L_{ext} $ unaffected and this fact allows us to keep $ L_{ext} $ at the same value of Sec. \ref{The problem}. In order to take advantage of this fact while relying on the assumption of unaffected $ D_R$ no more, it is useful to rewrite \eqref{QM} in the form $Q_{\max} \propto \frac{1}{L_{ext}}\frac{1}{V^{1.8}} (\frac{V^2}{D_R})^{1.8}$; a result which, according to our result $ P_{I = I_{peak}} = \frac{V^2}{D_R} $ above, leads to $Q_{\max} \propto \frac{1}{L_{ext}}\left[\frac{P_{I = I_{peak}}}{V}\right]^{1.8} $. To understand how $ Q_{\max} $ depends on $ V $ we need therefore to understand how $ P_{I = I_{peak}} $ depends on $ V $.\\ 

To this purpose, we must take into account that suppression of filamentation affects $ P $. We have shown in Sec. \ref{A model for saturation} that $ P_{I = I_{peak}} = \left[\frac{dW_{sh}}{dt}\right]_{I = I_{peak}} $. Now, as far as we are concerned with the dependence of $ Q_{\max} $ on $ V $ we may say that $ \left[\frac{dW_{sh}}{dt}\right]_{I = I_{peak}} $ is linear in $ \left[W_{sh}\right]_{I = I_{peak}} $. This statement makes sense because $ \left[\frac{dW_{sh}}{dt}\right]_{I = I_{peak}} \propto O ( \frac{\left[W_{sh}\right]_{I = I_{peak}}}{\tau_{LC}} )$. In turn, $ \left[W_{sh}\right]_{I = I_{peak}} $ grows with increasing $ E $ at least as $ \propto E^2 $ (again, see Sec. \ref{A model for saturation}). Finally, we invoke $ E = \frac{CV^2}{2} $. We line these results up and write $ P_{I = I_{peak}} = \left[\frac{dW_{sh}}{dt}\right]_{I = I_{peak}} \propto W_{sh} \propto E^2 \propto V^4 $. It follows that $ \frac{P_{I = I_{peak}}}{V} \propto V^3 $, hence $Q_{\max} \propto \frac{V^{5.4}}{L_{ext}} $ so that:\\ 
\begin{equation}
\label{comparisonnofilament}
\dfrac{Q_2}{Q_1} = \left(\dfrac{V_2}{V_1}\right)^{5.4}
\end{equation}\\
Let e.g. a PF with $ V_1 = $ 50 kV achieve $ Q_1 = Q_{record} = 3 \cdot 10^{-4}$. To achieve breakeven ($ Q_{2} = 1 $), we need $ V_2 = $ 224 kV. If $ Q_{record} $ is obtained at $ E = $ 0.5 MJ then breakeven corresponds to $ E = $ 10.04 MJ - a practically achievable value.\\ 

Suppression of filamentation makes breakeven more feasible because no more quasi-force-free magnetic fields mean more Lorenz force, and more Lorenz force means more kinetic energy available to the sheath for further compression into the pinch; this kinetic energy is $ \propto v^2 $ and increases therefore with $ V $ more rapidly than the dissipated power included in $ P \propto D_R \propto v $.\\

Remarkably, \eqref{comparisonnofilament} is even too pessimistic - and not just for our conservative assumption converning the scaling of $ \left[W_{sh}\right]_{I = I_{peak}} $ vs. $ E $. Our discussion, indeed, focuses on the electromagnetic side of sheath physics. As such it overlooks other negative effects of filamentation. As the advancing sheath moves axially, it meets the neutral gas. The mean free path for the process of resonant charge exchange is small enough for the range of gas operating pressures, so that gas ionization and its sweeping takes place at the front of the sheath. The efficiency of gas sweeping is one of the important conditions affecting the formation of a plasma pinch of suitable density and temperature. When it comes to gas sweeping a sheath made of filaments is of course less efficient than a smooth, axisymmetric sheath, as the neutral atoms can pass unscathed across the gaps between adjacent filaments \cite{SCHOLZ}: in the words of Ref. \cite{HAHN} \textit{a direct consequence of increased filamentation is a reduction in mass sweeping efficiency}. Suppression of filamentation can therefore improve gas sweeping and raise $ Y_{n} $ further. \\

Moreover, during rundown part of the current flows outside the sheath, mostly in the plasma generated during the ionization of residual working gas that has not been completely swept during detachment of the sheath from the insulator. As a result, only part of the current flows in the pinch, which becomes particularly evident as energy in the capacitor bank increases. Residual gas or partially ionized plasma remaining behind the sheath, caused by poor sweep efficiency, can lead to a decrease in dielectric strength of the space between the electrodes behind the sheath, and thus the appearance of an unwanted shunt effect \cite{NUKULIN}, which significantly reduces compression of plasma in the pinch. In megajoule systems, the value of the current flowing in the pinch decreases to even less than 50\% of the total current registered in the circuit \cite{SCHOLZ}. We may therefore reasonably assume that the better the gas sweeping, the smaller the fraction of current lost this way; and we have shown above that suppression of filamentation improves gas sweeping. \\ 

These conclusions bring to mind the working principle of FDEs \cite{NARDI2}: a FDE concentrates the interelectrode current on a thin sheath, starting from the onset of the discharge at the breech, and effectively eliminates diffuse distributions of current behind a high-density sheath. As a result, the fraction of the interelectrode current eventually flowing in the pinch grows \cite{NARDI}.

\section{How to suppress filaments}\label{How to suppress filaments} 
How to suppress filaments in the sheath? According to \cite{LERNERCONFINEMENT}, filamentation never occurs if the level of impurities exceeds the threshold hinted at in Sec. \ref{A model for saturation}; however, high impurity content may lead to excessive radiation losses, which in turn can prevent breakeven.\\

Experiments show mitigation of filamentation through some of the approaches listed in Sec. \ref{Introduction}. However, if filaments are the attractor of a relaxation process which drives the plasma sheath towards minimization of some quantity - as in Refs. \cite{AULUCKDESCRIPTION}, \cite{DIVITAHOTSPOT} and \cite{FERROFONTAN} - then their structure depends on the initial conditions of the discharge - which these approaches work upon - only weakly, and the same applies to the impact of filaments on $ Y_n$.\\

Remarkably, however, when reviewing some models of filamentation in the PF during rundown, it turns out that all these models suggest the same, natural way of suppressing filaments. In particular, each model provides us with its own sufficient condition for the prevention of an instability which leads to filamentation; even if these conditions are wildly different from each other, we are going to show how to satisfy all of them simultaneously. Accordingly, firstly we discuss some filamentation-triggering instabilities and highlight the corresponding sufficient conditions for stabilization, then we outline one method to meet all of them at once.\\

A so-called 'thermal instability' ('TI'), originally investigated in laser plasmas, may trigger filamentation \cite{FERROFONTAN}. Two small perturbations of electron temperature $ T_e $ and radial magnetic field $ B_r $ which depend on the azimuthal degree of freedom affect an unperturbed sheath where there is no radial magnetic field and both gradients of density and temperature point in the same (axial) direction. The baroclinic magnetic source term and the Leduc-Righi term in the induction equation for the time derivative of the radial magnetic field and the heat equation for time derivative of the temperature respectively make the two pertubations to reinforce each other. For given thermal conduction - with diffusion coefficient $\chi_{\bot} $ - across the perturbing radial magnetic field (i.e., in the azimuthal direction), the resulting instability is damped by resistive diffusion - with diffusion coefficient $ \frac{\eta_{\Omega \bot}}{\mu_0} $, $ \eta_{\Omega \bot} $ perpendicular resistivity. According to equations (14) and (19) of Ref. \cite{FERROFONTAN}, indeed, if TI occurs then $ \chi_{\bot} >  \frac{\eta_{\Omega \bot}}{\mu_0} $ i.e., $ \left( \frac{T_e (eV)}{2.21} \right)^4 \cdot \Lambda^{-2} \cdot \left( \frac{3.53 \cdot 10^{16}}{n_e (cm^{-3})}\right) > 1$, where $ \Lambda $ and $ n_e $ are the Coulomb logarithm and the electron density respectively. This result makes sense, as $ \eta_{\Omega \bot} \propto \Lambda T_e^{-\frac{3}{2}} $ and $ \chi_{\bot} \propto \frac{T_e}{\nu_{ee}} $ where  
$\nu_{ee} (s^{-1}) = 2.91 \cdot 10^{-6} \cdot \Lambda \cdot n_e (cm^{-3}) \cdot T (eV) ^{-\frac{3}{2}}$ 
%$\nu_{ee} \propto \Lambda n_e T_e ^{-\frac{3}{2}}$ 
\cite{NRL} is the electron-electron collision frequency. As a consequence, the hotter the sheath the more prone to filamentation. However, this result suggests that if we are able to lower $\chi_{\bot} $ we may suppress TI. We can lower $\chi_{\bot} $ by adding an unperturbed radial field $ B_{r0} $, so that $ \chi_{\bot} $ gets multiplied by $ \left[ 1 + \omega_{r}^2 \nu_{ei}^{-2} \right]^{-1}  < 1$ where $ \omega_{r} (rad \cdot s^{-1} $) $ = 1.76 \cdot 10^{11} B_{r}$ (T) \cite{NRL} is the electron cyclotron frequency related to the radial field. The contribution of the latter to the Lorenz force acting on the sheath vanishes because the electric current density too is radial. Intuitively, a sufficient condition for preventing TI is therefore: $ \left( \frac{T_e (eV)}{2.21} \right)^4 \cdot \Lambda^{-2} \cdot \left( \frac{3.53 \cdot 10^{16}}{n_e (cm^{-3})}\right) \cdot \left[ 1 + \omega_{r}^2 \nu_{ei}^{-2} \right]^{-1} < 1 $, 
%\[\left( \dfrac{T_e (eV)}{2.21} \right)^4 \cdot \dfrac{1}{\Lambda^{2}} \cdot \left( \dfrac{3.53 \cdot 10^{16}}{n_e (cm^{-3})}\right) \cdot \dfrac{1}{1 + \omega_{r}^2 \nu_{ei}^{-2}} < 1\]
i.e.:\\
\begin{equation}
\label{thermalinstabilitysuppression}
B_r \mbox{(T)} > B_{(TI)} \equiv 0.58 T_e (eV)^{-\frac{3}{2}} \sqrt{\Delta}  \quad \mbox{if} \quad \Delta > 0 \quad , B_{(TI)} = 0 \quad \mbox{otherwise}
\end{equation}\\
%\[\mbox{where }\Delta \equiv \left( \dfrac{T_e (eV)}{2.21} \right)^4 \left( \dfrac{n_e (cm^{-3})}{3.53 \cdot 10^{16}} \right) - \Lambda^{2}  \left( \dfrac{n_e (cm^{-3})}{3.53 \cdot 10^{16}} \right)^2\]\\
where $ \Delta \equiv \left( \frac{T_e (eV)}{2.21} \right)^4 \left( \frac{n_e (cm^{-3})}{3.53 \cdot 10^{16}} \right) - \Lambda^{2}  \left( \frac{n_e (cm^{-3})}{3.53 \cdot 10^{16}} \right)^2 $. Admittedly, \eqref{thermalinstabilitysuppression} is no rigorous result, as the model of Ref. \cite{FERROFONTAN} assumes $ B_{r0} = 0 $ so that $ B_r $ is just an estimate for the typical amplitude of a magnetic fluctuation around zero. According to Ref. \cite{FRUCHTMAN}, however, if we assume both $ B_{r0} \neq 0 $, $ \omega_{r} \gg \nu_{ei} $ and negligible dissipation then $ B_{r0} $ damps TI growth - and may even suppress it whenever the gradients of unperturbed density and temperature point in the same direction, as in the PF sheath. Accordingly, we still adopt \eqref{thermalinstabilitysuppression} as a (possibly over-restrictive) sufficient condition for suppression of TI-induced filamentation.\\

The perturbation of the magnetic field acts also on the current-carrying electrons themselves through Lorenz force. Small magnetic perturbations cause electrons to cluster into filaments, creating a feedback loop that increases local currents and enhances the magnetic field. Analysis of this 'current filamentation' (or 'Weibel') instability ('CFI') usually stems from the equations of motion of two electron beams streaming agains each other with opposite velocities with the same intensity $ v_{0}$; the only role of the much heavier ions is to ensure quasineutrality. In the plasma sheath during rundown, in contrast, electrons move radially at velocity $ v_{drift} $ from the cathode towards the anode; in the unperturbed, uniform sheath of thickness $ d_{i} $ we take as unperturbed state before CFI onset with $ v_{drift} = \frac{I}{2 \pi a \cdot d_{i}}\frac{1}{n_e e}$, $ e $ elementary charge. 
%For mathematical simplicity, let us assume a slab geometry (the same choice of Ref. \cite{FERROFONTAN}, by the way). 
Galileian invariance allows us to select the coordinate frame where one of the beams is at rest and we retrieve our sheath, provided that $ v_{0} = \frac{v_{drift}}{2} $. The wavevector of the excited mode is perpendicular to the current density and has intensity $ k \approx O \left( d_e^{-1}\right) $ where $ d_e = \sqrt{\frac{m_e}{m_i}} \cdot d_i$ is the collisionless electron skin depth. An externally applied magnetic field parallel to the current density - i.e., radial - can suppress the instability. In the PF sheath $ v_{0} \ll $ the speed $ c $ of light in vacuum. Then, a sufficient condition of CFI stability \cite{STOCKEM} is $ \omega_{r} > d_e^{-1}v_{drift} $, 
%\\ \[\omega_{r} > \dfrac{v_{drift}}{d_e} \]
i.e. (in deuterium):\\
\begin{equation}
\label{Weibelinstabilitysuppression}
B_r \mbox{(T)} > B_{(CFI)} \equiv 3.3 \cdot 10^{-3} \cdot I(MA) \cdot a(m)^{-1}
\end{equation}\\
provided that the unperturbed azimuthal magnetic field $ B_{\vartheta 0}$ vanishes. However, it is precisely $ B_{\vartheta 0}$ which provides the Lorenz force pushing the sheath of plasma. Admittedly, if $ B_{\vartheta 0} \neq 0 $ then the magnetic field lines cannot be exactly parallel to the electric current density. It turns out that perfect stabilization is impossible in the general case. In the particular case of the plasma sheath during rundown, $ k v_{0}d_e c^{-1} \ll 1 $ and $ \omega_{b} d_e c^{-1} \ll 1 $ ($ \omega_{b} $ is the electron cyclotron frequency related to the total magnetic field with intensity $ \sqrt{B_{r0}^2+B_{\vartheta 0}^2} $); in this limit, Fig. 1 of Ref. \cite{BRET} shows that the growth rate of CFI vanishes. If $ B_r = 0 $, however, no stabilization is ever possible; accordingly, we stick to \eqref{Weibelinstabilitysuppression} as for the sufficient condition for the suppression of CFI-induced filamentation.\\

According to Ref. \cite{TSYBENKOERISKIN}, PF filament generation is associated with the corrugation instability of a rarefaction shock wave ('RSW'). A central tenet of the model is the assumption that the electric current density of the filament satisfies London equation with penetration depth $ d_e $. Should the magnetic field of the filament be exactly radial, i.e. parallel to the filament axis, then it would be described by equation (8.7) of Ref.  \cite{DIVITAHOTSPOT} and its typical amplitude $ B_f $ would satisfy the relationship $ B_f  = \left( 2 \pi \right)^{-1} d_e^{-2} \Phi $ where $ \Phi = 2 \pi \left( m_e \vert e \vert^{-1} \right) \left( T_e \vert e \vert^{-1} B_f^{-1} \right)$ is the flux of $ B_f $, so that $ B_f = \sqrt{\mu_0 p_e} $ with $ p_e = n_e T_e $ electron pressure; in this case, application of an external, radial field $ B_r > B_f $ would be enough to suppress filamentation. This condition, however, is likely to be overpessimistic. If filaments are quasi-force-free, indeed, their magnetic field has not only the axial component $ B_f $, but also a component $ B_{\bot} \approx B_f $ perpendicular to the filament axis, and the corresponding contributions $ \propto  B_f^2 $ and $ \propto  B_{\bot}^2 $ to $ W_{sh} $ and the sheath inductance are $ \approx $ equal. All other things being equal, accordingly, we expect the actual value of $ B_f $ in a real filament to be $ \approx \frac{1}{\sqrt{2}} $ the value $ \sqrt{\mu_0 p_e} $ obtained above for a purely radial field. Thus, we expect suppression of RSW-induced filamentation if $ B_r > \sqrt{\frac{\mu_0 p_e}{2}} $, i.e.: 
%, the value   - and both . It follows that a fulament with given energy   it would be  is the magnetic flux of the filament (YYYYY DA RIFARE DA QUI AL PROSSIMO YYYYY  According to Ref. \cite{DIVITAHOTSPOT}, This postulate, however, is just a consequence of the result of Ref. \cite{DIVITAHOTSPOT}, according to which the filamentary structure of the sheath is a relaxed state corresponding to a minimum of a Ginzburg-Landau functional, - in analogy with type-II superconductors - where the squared absolute value of the order parameter is $ \propto n_e $. The analogy with superconductors suggests that a minimum magnetic field $ \approx \vert a_{GL} \vert \sqrt{\frac{2 \mu_0}{b_{GL}}} $ parallel to the filament direction exists which destroys the filamentary structure, where $ a_{GL} $ and $ \frac{b_{GL}}{2} $ are the first two coefficients of the development of the functional in powers of the order parameter. Their values are $ a_{GL} = - \frac{p_0e}{\rho_0e} $ and $ \frac{b_{GL}}{2} = \frac{p_0e}{2 \rho_0e ^2} $, where $ p_{0e} $ and $ \rho_{0e} $ are the unperturbed pressure and mass density of electrons. Thus, we expect suppression of RSW-induced filamentation if:\\
\begin{equation}
\label{RSWsuppression}
B_r \mbox{(T)} > B_{(RSW)} \equiv 3.15 \cdot 10^{-13} \sqrt{n_e (m^{-3}) T_e (eV)} 
\end{equation}\\
Both TI, CFI and RSW take into account no perturbation of density; moreover, their typical length is $ d_e $, rather than $ d_i $. But we know that vortex-filaments \cite{SOTO} in the plasma sheath during rundown involve the macroscopic velocity, which is basically related to the motion of ions. When it comes to the macroscopic velocity in a magnetized plasma accelerated towards a less dense fluid - like the plasma sheath pushed by Lorenz force in the PF - the instability of choice is the 'magnetic Rayleigh-Taylor instability' ('MRTI'). If in a set of $ N_f $ filaments which carries a current $ I $ we describe one radial filament which an electric current $ I_{f} = I \cdot N_f^{-1} $ flows across as a dynamic Z-pinch with radius $ R_f $ ($ = d_i $ \cite{AULUCKDESCRIPTION}) and apply a radial magnetic field $ B_{r} $, then a sufficient condition for the suppression of MRTI-induced filamentation is \cite{BUDKO}: \\
\begin{equation}
\label{magneticRayleighTaylorinstabilitysuppression}
B_{r} \mbox{(T)} \geq B_{(MRTI)} \equiv K \cdot I_{f} \mbox{(MA)} \cdot R_f \mbox{(cm)}^{-1} 
\end{equation}\\ 
where $ 1 \leq K \leq 3$ is a dimensionless constant. Experiments show mitigation of MRTI with the help of an applied 0.4 T axial magnetic field in a 300 kA Z-pinch with radius 1 cm - see Fig. 1.f of Ref. \cite{MIKITCHUK}. Then, we take $ K = 1.33$.\\

\begin{figure} 
\includegraphics[scale=0.4]{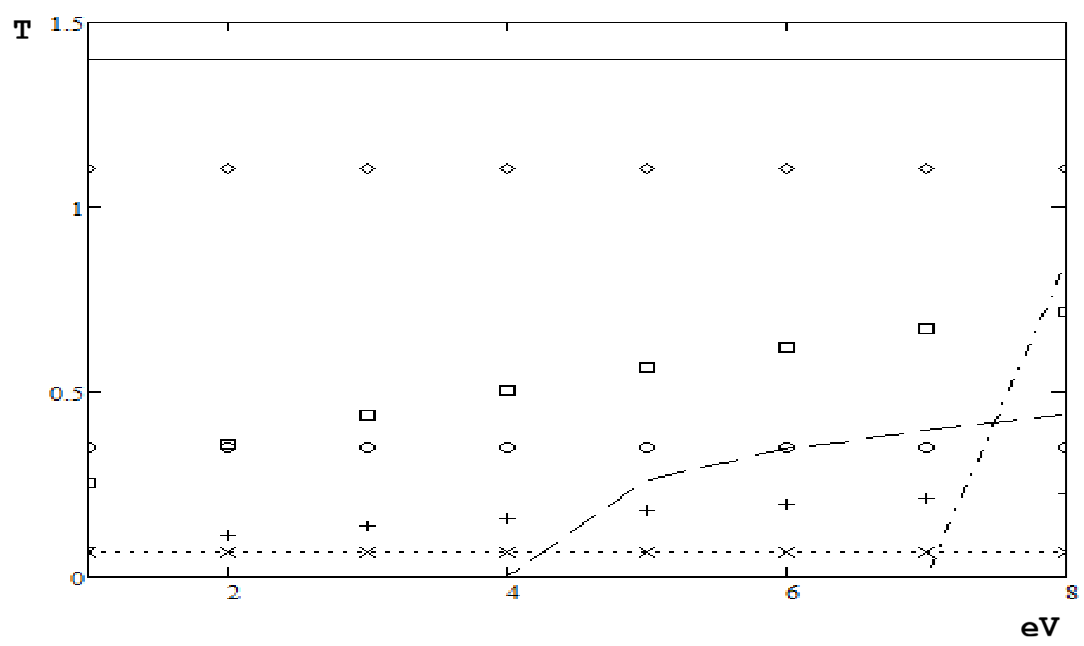}
\caption{\textit{Minimum values of the radial magnetic field (T) required for the suppression of filamentation-inducing instabilities vs. $ T_e $ (eV) for $ n_e = n_m $ and $ n_e = 10 \cdot n_m $. (Dashed line): $ B_{(TI)} $, $ n_e = n_m$. (Dash-dotted line): $ B_{(TI)} $, $ n_e = 10 \cdot n_m$. (XXX): $ B_{(CFI)} $, $ n_e = n_m$. (...): $ B_{(CFI)} $, $ n_e = 10 \cdot n_m$. (+++): $ B_{(RSW)} $, $ n_e = n_m$. ($ \square \square \square $): $ B_{(RSW)} $, $ n_e = 10 \cdot n_m$. (ooo): $ B_{(MRTI)} $, $ n_e = n_m$. ($ \lozenge \lozenge \lozenge $): $ B_{(MRTI)} $, $ n_e = 10 \cdot n_m$. Continuous horizontal line: $ B_{PM} $.}}
\label{B} 
\end{figure}

According to \eqref{thermalinstabilitysuppression}, \eqref{Weibelinstabilitysuppression}, \eqref{RSWsuppression} and \eqref{magneticRayleighTaylorinstabilitysuppression}, superimposition of a radial magnetic field larger than some threshold in the interelectrode region of the PF where rundown occurs suppresses four distinct filamentation processes, namely TI, CFI, RSW and MRTI. It is useful to compare the corresponding thresholds $ B_{(TI)} $, $ B_{(CFI)} $, $ B_{(RSW)} $ and $ B_{(MRTI)} $ with the intensity $ B_{PM} $ of the residual field of a permanent magnet in typical ranges $ 1 \leq T_e (eV) \leq 8 $ and $ n_m \leq n_e (cm^{-3}) \leq 10 \cdot n_m $ for $ T_e $ and $ n_e $, where $ n_m \equiv 6.44 \cdot 10^{16}$ cm$ ^{-3}$ corresponds to an initial 1 Torr pressure of the working gas (deuterium) in the interelectrode space. To fix ideas, we focus on the case $ I = 1 $ MA, $ a = 5 $ cm, $ R_f = \delta_i $ and $ N_f = 30$; we take also $ B_{PM} = 1.4$ T, a value obtained e.g. with commercially available, NdFeB magnets. In all cases, it turns out that the field produced by a permanent magnet is able to stabilize the instability, hence to prevent filamentation - see Fig.~\ref{B}. 
\\

\begin{figure} 
\includegraphics[scale=0.3]{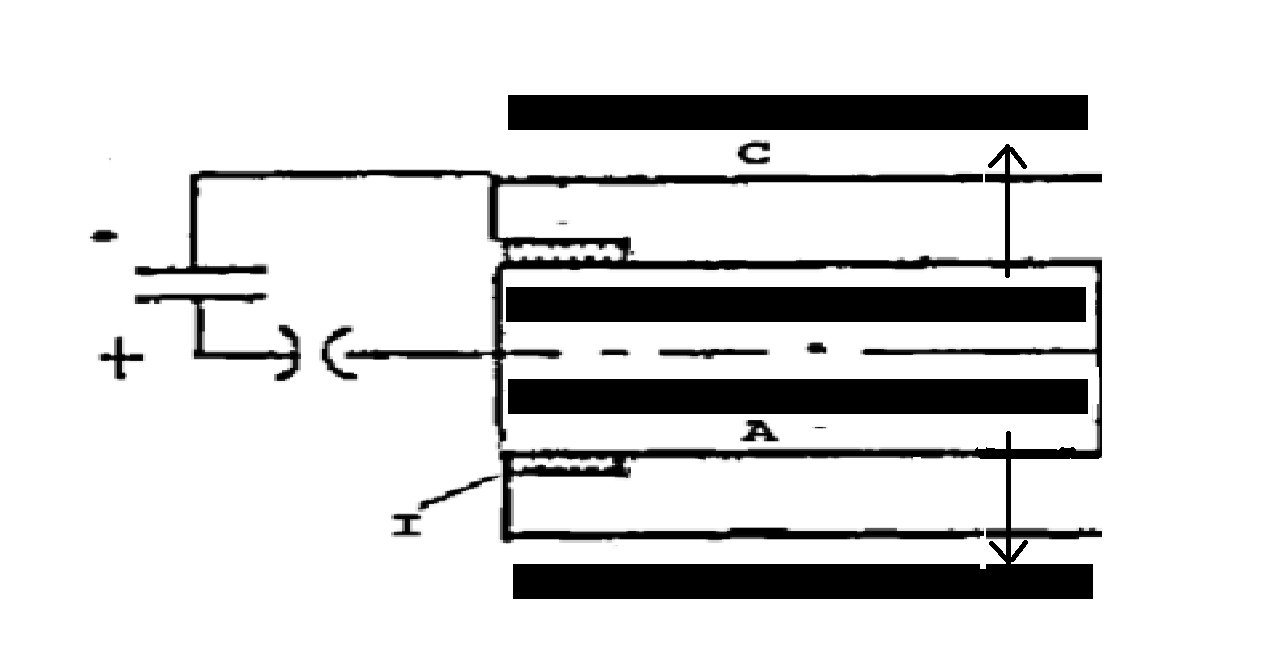}
\caption{\textit{Possible location of the sources (the black rectangles) of the radial magnetic field $ B_r $ (the arrows) in the PF of Fig.~\ref{PF} (cross section).}}
\label{HALL} 
\end{figure}

In order to produce the radial magnetic field, the lay-out of the permanent magnets is likely to be similar to the lay-out of the sources of the radial magnetic field in the cylindrical geometry of a Hall thruster \cite{GOEBEL} for space propulsion - see Fig.~\ref{HALL}. We recall that both the electrodes and the insulator of a PF are typically made of non-magnetic materials like e.g. copper, stainless steel, beryllium or alumina; they are therefore unlikely to bend the magnetic field lines.
One could argue that the radial magnetic field produced this way is not uniform; its intensity is actually decreasing as the radial distance from the anode surface increases. But then, we recall again that filaments are well distinguishable precisely near the anode surface, i.e. in the region where the intensity of the stabilizing $ B_{PM} $ is maximum.\\

Admittedly, our discussion is far from embracing all possible configurations. Then, it is perfectly possible that the actual value of the field required by the suppression of filamentation exceeds the maximum field produced by a permanent magnet. In this case, suitably arranged current-carrying high temperature superconductors ('HTS') may generate a multiTesla field with limited power consumption, thus replacing the permanent magnet. Maximum HTS field values are $ \approx 20 $ T and $ \approx 45 $ T in continuous and pulsed mode respectively. For dimensional reasons, the current flowing across a HTS which is located inside a PF anode with radius $ a $ and which produces a field $ B_r $ is $ \approx O \left( \frac{2 \pi a B_r}{\mu_0} \right) $, and the corresponding current density is $ \frac{1}{\pi a^2} $ this value, i.e. $ \approx O \left( \frac{2 B_r}{a \mu_0} \right)$. For $ B_r = 1.4 $ T and $ a = 0.115 $ m (the anode radius of PF1000) we have $ 19 $ $ \frac{A}{mm^2} $, safely below the maximum value allowable for the superconducting state to avoid quenching (800 $ \frac{A}{mm^2} $ at 4.2 K). Admittedly, the magnetic energy stored in $ B_r $ is not negligible (1.4 T in a volume with $ a = 0.115$ m, $ b = 0.2 $ m and $ h = $ 0.6 m like in PF1000 correspond to 39 kJ). But then, this amount of energy is to be stored once for all and can suppress filamentation in many discharges in succession. Finally, $ B_r $-related mechanical stresses on the anode are $ \approx O( \frac{B_r^2}{\mu_0} \cdot a \cdot h ) \approx 10^5 $ N and are definitely not negligible; their impact, however, is likely to be less negative than the repeated, pulsed stresses of the same order of magnitude ($ \approx 10^5 $ N) \cite{MIKHAILOV} due to the repeated multi-MA discharges.\\  

Not all filamentation-inducing instabilities are suppressed with the help of a superimposed field, though. Speaking of $ B_{\vartheta 0} $, for example, we recall that filaments are radial \cite{BOSTICK}. It is therefore only natural to ask whether they can be the final outcome of an instability propagating along the azimuthal direction, i.e. parallel to the existing magnetic field $ B_{\vartheta 0} $. This is the case of the so-called 'electrothermal instability' ('ETI') \cite{AreviewofthedenseZpinchHAINES}. The essential mechanism is that a spatially varying perturbation in $ T_e $ orthogonal to the direction of the current density will lead to perturbed current density and Joule heating. This is due to the T$ ^{\frac{3}{2}} $ power law for the electrical conductivity which will lead to perturbed Joule heating that will enhance the perturbed temperature. 
%ETI occurs if $ T_e $ (eV) $ < 4.11 \cdot 10^{-6} \cdot  n_e $(m$^{-3}$)$^{\frac{1}{4}} \cdot \Lambda^{\frac{1}{2}} $ and, simultaneously, 
A necessary condition for the occurrence of ETI is $ \frac{T_e}{T_i} > 1.32$ with $ T_i $ ion temperature. A break-up of the current into filaments is expected as a consequence of the instability. If the energy lost by the plasma due to radiation is $ \propto T_e ^\zeta $ with $\zeta < 1 $ ($ \zeta > 1 $) then radiation losses enhance (suppress) the growth of ETI \cite{ORESHKIN}; accordingly, if Bremsstrahlung is the only radiation process ($ \zeta = \frac{1}{2} $) then radiation leads to no stabilization. In this case the fastest growing mode has %wavelength $ \lambda $(m) $= 0.56 \sqrt{A} \cdot 10^{20} \cdot T_e (eV)^2 n_e(m^{-3})^{-1} \Lambda ^{-1}$ and 
a growth rate $ \gamma_{(ETI)} (s^{-1}) = 1.77 \cdot 10^{-12} \cdot \frac{m_e}{m_i} \cdot n_e(m^{-3}) \cdot \Lambda \cdot T_e (eV)^{-\frac{3}{2}}$ with $ A $, $ m_e $ and $ m_i $ atomic mass (=2 in deuterium), electron mass and ion mass respectively. 
%Of course, existence of filaments inside the sheath requires $ \lambda \leq 2 \pi a $. Moreover, there is enough time during rundown to see a significant development of ETI if $ \gamma > \frac{h}{v} $. Finally, in 
In order to allow $ \frac{T_e}{T_i} > 1.32$ during rundown the typical time-scale $ \frac{1}{\gamma_{(ETI)}} $ of ETI growth must be shorter than the thermal equilibration time $ \tau_{eq} \propto A T_e^{\frac{3}{2}} \Lambda^{-1} n_i^{-1}$ 
%, i.e. $ \gamma > \frac{1}{\tau _{eq}} $(s$^{-1}$)$ = 3.2 \cdot 10^{-9} \cdot A^{-1} \cdot \Lambda \cdot n_i (cm^{-3}) \cdot T_{e} (eV) ^{-\frac{3}{2}}$ 
($ n_i = n_e $ ion density). And in order to allow time for the instability to grow $ \frac{1}{\gamma_{(ETI)}} $ must also be shorter than $ \approx O ( \tau_{LC} ) $. Thus, a sufficient condition for suppression of ETI-induced filamentation is:\\
%\begin{equation}
%\label{electrothermalinstabilitysuppression}
%T_e (eV) \geq 4.11 \cdot 10^{-6} n_e \left(m^{-3} \right)^{\frac{1}{4}} \Lambda^{\frac{1}{2}} \quad  
%\mbox{or} \quad \lambda > 2 \pi a \quad \mbox{or} \quad \gamma \leq \max \left( \dfrac{h}{v}  ,  \dfrac{1}{\tau _{eq}} \right)  
%\end{equation}
\begin{equation}
\label{electrothermalinstabilitysuppression}
\gamma_{(ETI)} \leq \max \left( \tau _{eq}^{-1} , \tau_{LC}^{-1} \right) 
\end{equation}\\
Typically, $ \tau_{LC} = 10^{-6} $ s and $ \frac{1}{\tau _{eq}} $(s$^{-1}$)$ = 3.2 \cdot 10^{-9} \cdot A^{-1} \cdot \Lambda \cdot n_i (cm^{-3}) \cdot T_{e} (eV) ^{-\frac{3}{2}} $ \cite{NRL}. Then,  \eqref{electrothermalinstabilitysuppression} is always satisfied in the ranges of $ n_e $ and $ T_e $ considered above. Accordingly, we refer to ETI no more in the following.
\section{Conclusions} \label{Conclusions}
It was observed early in Plasma Focus ('PF') research that $ Y_n \propto E^2$ in a pure deuterium plasma, where $ Y_n $ and $ E $ are the neutron yield and the capacitor storage energy respectively. Such scaling gave hopes of possible development as a fusion energy source, because it implies that the fusion gain $ Q \propto \frac{Y_n}{E} $ - increases linearly with increasing $ E $. Devices were scaled up to higher $ E $. It was then observed that the scaling deteriorated, with $ Y_n $ not increasing as much as suggested by the $ E^2 $ scaling. Deterioration of this scaling became apparent at $ E $ approached 0.5 MJ. Correspondingly, $ Q $ achieves a maximum ('saturation'); it decreases as $ E $ increases further. To date, there is still no convincing explanation of neutron saturation; both careful selection of materials (e.g. concerning the insulator sleeve separating the electrodes), addition of suitable field distortion elements, iterated optimization of the geometry of electrodes, and deeper insight in the pinch physics ensured no real breakthrough. Historically, both lack of understanding of the fundamental physics of saturation, competition of better-performing tokamaks and enduring pessimism concerning beam-plasma fusion - which plays a crucial role in PFs - led to shutdown of major fusion-oriented PF back in the Eighties. To date, there are no more PFs aimed at fusion power production with $E \geq $ 1 MJ; they are rather seen as intense, pulsed neutron sources for material testing \cite{GOYONMJOLNIR}.\\

Not surprisingly, today we can find just four proposals for achieving breakeven in PFs. The first one, and by far the more developed, is due to LPPFusion \cite{LERNER}. It aims at creating tiny plasmoids inside the pinch. It is believed that the magnetic field in these plasmoids is so large ($ \geq 10^4 $ T) that quantum mechanical effects alter the physics of collisional exchange between plasma particles, even paving the road towards aneutronic, p-$ ^{11} $B fusion. In spite of the considerable experimental results achieved to date by LLPFusion, however, the actual relevance of these quantum mechanical effects to the breakeven in PFs is still disputed \cite{BAALRUD} \cite{DIVITAPB}.\\

Alternatively, the sophisticated model referred to in Ref. \cite{AULUCKSATURATION} stems from the results of Ref. \cite{GRATTONVARGAS}, takes explicitely into account the curved shape of the sheath in the rundown phase and leads to a spectacular prediction: a 60\% reduction of the radial size of the insulator sleeve can multiply $ Y_{n} $ by 200. However, the model relies on the rather overoptimistic assumption that the capacitor bank is fully discharged at the time when the plasma arrives at the anode centre.\\ 

Finally, the widely accepted, macroscopic, simple but robust Lee's model \cite{LEE2014} allows us to identify the change in time of the inductance $ L $ in the rundown phase ('dynamic resistance' $ D_R $) as the physical quantity which is to be accounted for the saturation of $ Y_n $ \cite{LEE}. In a nutshell: as the plasma moves during rundown, its inductance grows; in turn, this very growth acts as an electric resistance, which hinders further energy supply from the condenser bank to the plasma. As we raise the energy $ E = \frac{CV^2}{2}$ available to the plasma by raising the capacitance $ C $ of the capacitor bank which feeds the PF at given voltage $ V $, indeed, the impedance $ \sqrt{\frac{L}{C}} $ becomes negligible when compared to $ D_R $. The finite value of $ D_R $ puts therefore an upper bound on the peak value $ I_{peak} $ of the plasma current, which $ Y_n $ is a strongly increasing function of \cite{LEE}.\\

A somehow related analysis is presented in Ref. \cite{NUKULIN}, where it is shown that raising $ C $ slows the discharge of the capacitor bank and forces us to lenghten the anode in order to ensure the synchronism between the end of the rundown and the instant of maximum current, as required by maximization of $ Y_n $. The longer the electrodes, the larger the overall PF inductance, the lower $ I_{peak} $, the lower $ Y_n $, and saturation follows.\\

Both Refs. \cite{LEE} and \cite{NUKULIN}, however, lead to the same results for the purposes of our discussion. Indeed, both agree on how to overcome saturation: just raise $ E $ while leaving $ C $ unchanged, i.e. by raising $ V $. Unfortunately, it turns out that achieving breakeven in a DT plasma this way requires values of $ V $ and $ E $ in the multi-MV and multi-GJ range respectively - easier said than done.\\

As for achieving breakeven in a PF, admittedly, the treatments of both Refs. \cite{LEE}, \cite{NUKULIN} and \cite{AULUCKSATURATION} are still in their infancy. In comparison with the ambitious goals of LLPFusion \cite{LERNER}, however, both enjoy the advantage of well-established physics; moreover, the control parameters they claim to be crucial for breakeven are both macroscopic (the shape of the insulator, the voltage of the capacitor bank) and easy to control - in principle at least. Finally, both models are concerned with the physics of the sheath during rundown, and are therefore unaffected by the considerable difficulties in modelling and observing the physics of the pinch. They share a common, robust point of view: everything spoiling the transfer of energy from the condenser bank to the plasma is eventually going to reduce the energy available to the pinch, hence the number of nuclear reactions per shot - regardless of the subtleties of the detailed pinch physics. Given the extreme richness of the latter \cite{AULUCKSADOWSKIMIKLASZEWSKI}, this is a considerable advantage when it comes to the problem of achieving breakeven.\\

Being unaffected by overoptimistic assumptions, the simple model of Ref. \cite{LEE} deserves particular attention. (As for the results our discussion stems from, Ref. \cite{NUKULIN} and Ref. \cite{LEE} lead basically to the same conclusions; we keep therefore on referring ourselves to the latter in the following). Its main limit is that the value of $ D_R $ - the fundamental quantity of the model - is to be given as an input: the model provides no information about it. As such, $ D_R $ can only be derived from observations which are still the outcome of experiments performed on PFs quite far from breakeven. Two equations - the balance of energy and of electric charge, in the form of Kirchoff's circuit laws - are available for three quantities, namely $ I_{peak} $ (which $ Y_n $ depends upon), $ D_R $ and the dissipated power (which affects the energy balance). To fill the gap, a third equation is required. To this purpose, we need a model of the saturation process (whose effect is embedded in $ D_R $).\\

The model of Ref. \cite{DIVITA1993} describes saturation as the consequence of the spontaneous growth of a 3D structure of radial, quasi-force-free vortex-current filaments \cite{SOTO} \cite{AULUCKDESCRIPTION} inside the sheath in the rundown phase. The words 'quasi-force-free' mean 'with electric current density almost parallel to the magnetic field'. The words 'vortex-current filament' mean 'well-localized concentrations of both electric current density and fluid vorticity'. Spontaneous filamentation of the distribution of electric current inside the plasma raises inductance further, thus contributing to $ D_R $. The third equation provided by the model to fill the gap is the balance of magnetic helicity. (This balance reduces to an identity in the 2D case).\\

Filaments in the sheath have been observed under different experimental conditions \cite{BOSTICK} \cite{AULUCKFILAMENTATION}. The idea that the sheath in the rundown phase is made of quasi force-free filaments agrees with the observations reviewed in Ref. \cite{WHBOSTICK} and fits also the description of the evolution of the sheath as a succession of relaxed states of a Hall magnetofluid \cite{AULUCKDESCRIPTION}.\\

It has been shown that relaxation to such states minimizes the dissipated power in the plasma, provided that the value of its Hartmann number $ Ha $ is not too large \cite{DIVITAHOTSPOT} \cite{DIVITAHARTMANN}, i.e. when the viscous force is not negligible when compared with the Lorenz force. In this framework, a simple explanation is available \cite{LERNERCONFINEMENT}: while a perfectly smooth, 2D sheath may sweep all the neutral gas in front of itself as it moves along the electrodes in the rundown phase, filamentation reduces the efficiency of sweeping and reduces the friction between the plasma and the neutral gas, as most of the latter can pass through the filaments from the region in front of the sheath towards the region behind it. \\

Popular models of the sheath are 2D and neglect filaments \cite{AULUCKSATURATION} \cite{LEE2014}. In contrast, the 3D nature of filaments allows them to store a significant amount $ W_{sh} $ of magnetic energy at the expense of the PF capacitor bank. Being their magnetic fields quasi force-free, moreover, they exert negligible Lorenz force on the sheath. Correspondingly, $ W_{sh} $ is wasted from the point of view of the mechanical work done by the external world upon the plasma. Then, the maximum amount of energy available to the compression of the pinch on front of the anode - where nuclear reactions occur - is no more $ \propto E $, but $ \propto \left( E - W_{sh} \right)$; and since it turns out \cite{DIVITA1993} that $ W_{sh} $ itself increases with increasing $ E $ at least as $ E^2 $, saturation follows at large enough values of $ E $. The results of Ref. \cite{DIVITA1993} agree with observations concerning saturation in the PF PF360.\\ 

As filamentation induces saturation, prevention of the latter requires suppression of the former - at least near the anode surface, i.e. where the Lorenz force is stronger and our effort to save mechanical work is likely to be more effective. Now, it is precisely near the anode surface that the images of the sheath show that filaments are particularly well distinguishable \cite{BOSTICK}. Even if the popular squirrel-cage geometry of the PF cathode  made of many rods seems uniquely fit to produce filaments in the sheath during rundown one filament per rod, in order to improve the impact of our efforts on the Lorenz force it is near the anode surface where we want to suppress filamentation as much as possible.\\

Filamentation reduces dissipation \cite{LERNERCONFINEMENT}. Then, its suppression raises dissipation. The negative impact of this increased dissipation on the PF performances, however, is overcompensated by the increase of the mechanical energy available to the sheath before the formation of the pinch, as no more energy is wasted in quasi force-free fields. All other things being equal, actually, the smaller the role of quasi-force-free fields, the larger the Lorenz force exerted on the plasma sheath during rundown, the larger both the sheath velocity $ v $, $ D_R \propto v $, the dissipated power $ \propto D_R $ and the sheath kinetic energy $ \propto v^2 $. Note that the latter energy grows more rapidly than the dissipated power; and so does also the amount of mechanical energy available to the compression of the pinch in front of the anode, after the rundown.  \\

Actually, it is true that we can achieve breakeven just by raising $ V $, as suggested by Refs. \cite{LEE} and \cite{NUKULIN}. In practice, however, this is only feasible if filamentation is suppressed. This suppression prevents energy waste, then allows us to achieve larger values of $ Q $ with the same $ V $ - or, equivalently, the same value of $ Q $ with lower $ V $. In particular, it turns out that that suppression of filamentation prevents saturation, multiplies the PF drive parameter by a factor $ 3 $ at least and allows breakeven in a 224 kV, 10 MJ PF working with 50\%deuterium-50\%tritium as the working gas. \\

It remains to be seen how to suppress filamentation in practice. Raising the impurity content does the job \cite{LERNERCONFINEMENT} but is obviously out of the question lest radiation losses are too large. Unfortunately, morever, if filamentation is really a particular case of a general trend towards minimization of dissipation  \cite{LERNERCONFINEMENT} \cite{DIVITAHOTSPOT} \cite{FERROFONTAN} then most strategies listed above are likely to be scarcely successful, as they act on the initial conditions only of the rundown - i.e., near the insulator sleeve. However, filaments are radial; thus, physical intuition suggests a radial magnetic field $ B_r $ can stabilize the instabilities leading to their formation, thus preventing filamentation.\\

A review of well-known filamentation-triggering instabilities - the 'thermal instability' \cite{FERROFONTAN}, the Weibel instability \cite{STOCKEM} \cite{BRET}, the corrugation instability \cite{TSYBENKOERISKIN}, the magnetic Rayleigh-Taylor instability \cite{BUDKO} and the electrothermal instability \cite{AreviewofthedenseZpinchHAINES} - confirms this suggestion and shows that the intensity required of the radial stabilizing fields in a range of electron density and temperature which is relevant to the PF sheath during rundown is not too far from 1.4 T: basically, the residual field of commercially available NdFeB permanent magnets. Should higher fields be required, their generation seems feasible with the help of high temperature superconductors ('HTS'). According to a preliminary estimate, the required current density in a HTS which produces the stabilizing field in the room available in a PF is too small to trigger unwanted quenching of the superconductor. Given the cylindrical symmetry of the PF, generation of a radial field is possible if the location of these magnets resembles the lay-out of the sources of magnetic field in a cylindrical SPT thruster for space propulsion, where radial fields are everyday's working tools \cite{GOEBEL}.\\

To the best of author's knowledge, no experiment has ever been performed on a PF near saturation where a radial magnetic field has been deliberately superimposed on the cylindrical region between the electrodes - yet. We line our results up and predict that a 224 kV, 10 MJ PF working with 50\%deuterium-50\%tritium and equipped with a radial magnetic field $ \approx $ 1.4 T achieves breakeven.\\

Admittedly, our discussion is qualitative - or, to be more accurate, it is not more rigorous than the original Lee's model of Refs. \cite{LEE} and \cite{LEE2014} it stems from. Moreover, it lacks a self-consistent treatment of electric current density and fluid vorticity on an equal footing when it comes to describe filaments. We focussed rather on the electromagnetic degrees of freedom - and invoked accordingly the MHD balance of magnetic helicity rather than the balance of generalized helicity of Hall MHD \cite{TURNER} - because we focussed on the smallness of the magnetic force in order to explain saturation. Moreover, our treament of radiation - limited to Bremsstrahlung - is definitely an unrealistic oversimplification. Furthermore, and unlike Ref.\cite{AULUCKSATURATION}, we neglected the canting of the sheath. Finally, the radial magnetic field and the axial velocity produce an azimuthal electric field, which has the same order of magnitude of the radial electric field between the electrodes. As a result, we expect the axis of the filaments to be bent in a spiral-like shape. A discussion of the consequences lies beyond the limits of the present work. A fully self-consistent treatment is the topic of future work.\\

But the robustness and the vast popularity of Lee's model, together with our use of well-established scaling laws \cite{PPCFSOTO}, ensure the generality of our main result: filament suppression during the rundown is a game changer, as far as the fusion gain is concerned. Moreover, the model relies on no hard-to-verify assumption concerning the complicated physics of the pinch, because it is only concerned with what happens in the rundown phase. Finally, the crucial assumption of the quasi force-free nature of filaments is a decade-old tool of PF research \cite{BOSTICKORIGINAL} \cite{WHBOSTICK} \cite{SESTERO}; the real novelty of the present work lies rather in working together with this assumption and with Lee's model both to discuss the connection between filamentation and saturation and to show the advantage of suppressing the former in order to remove the latter and achieve breakeven. \\

Some final remarks. As for controlled nuclear fusion, the remarkable advancement in the fusion triple product and the progress of technology lead to more and more optimistic expectations. In this regard, it is worthwhile to quote the conclusions of Ref. \cite{TAKEDA} verbatim: \textit{according to the collective remarks by scientists, the popular phrase 'fusion is always 30 years away' is proven wrong, technically speaking. To be precise, we should now say 'fusion was said to be 19.3 years away 30 years ago; it was 28.3 years away 20 years ago; 27.8 years away 10
years ago'. And now, scientists believe fusion energy is only 17.8 years away.} The last sentence of this work is: \textit{So there is a progress, and it is accelerating toward the realization of this ultimate clean energy.} On the other hand, however, skepticism is still alive and kicking. Sheer complexity of fusion power plants (in both magnetic and inertial confinement), huge capital expenditure and limited learning rates when it comes to cost reduction may possibly render fusion uncompetitive with alternative low-carbon power sources, such as nuclear fission, hydropower and geothermal \cite{TANGNOLL}.\\

In the author's opinion, we would better leave no stone unturned, if only out of prevention of further disappointment after decades of research with mixed results. Neutron saturation led to shutdown of major PF experiments forty years ago, when the Promised Land of breakeven was already in sight. When it comes to PFs, a noteworthy fact is that even modest resources lead to significant results. For example, the entire LPPFusion device fits within a small room 4 meters on a side; LPPFusion constructed it for a cost $ \approx $ \$500.000, and raised just 9 million \$ in private investment after the end of public funding in 2001 to date. Should we be really able to overcome saturation in a PF, then the compactness and the low cost of a PF reactor would be competitive beyond imagination in the fusion landscape.\\

If confirmed by experiments, the relevance of high-$ Q $ PF operation extends also to hybrid fission-fusion reactors: at breakeven with 50\% - 50\% tritium, a PF with 10 MJ energy condenser bank - like in the case discussed in Sec. \ref{From suppression of filaments to breakeven} - and with 1 Hz repetition rate - hence a production of $ \approx 4 \cdot 10^{18} $ neutrons per second - allows production of 200 MW thermal power in a volume $ \approx 20 $ m$ ^{3} $. An outstanding feature of the design is that no active components are necessary within the reactor containment area, all the hybrid system control being ensured by the fusion component of the reactor \cite{ZOITA}. This is likely to be the reason why some researchers seriously start to reconsider the past choice of shutting down research on PF-based fusion power production \cite{SOTOPRESENTATION} \cite{SADATKIAI}.

\bibliography{sn-bibliography}% common bib file

%% BioMed_Central_Bib_Style_v1.01

\begin{thebibliography}{75}
% BibTex style file: bmc-mathphys.bst (version 2.1), 2014-07-24
\ifx \bisbn   \undefined \def \bisbn  #1{ISBN #1}\fi
\ifx \binits  \undefined \def \binits#1{#1}\fi
\ifx \bauthor  \undefined \def \bauthor#1{#1}\fi
\ifx \batitle  \undefined \def \batitle#1{#1}\fi
\ifx \bjtitle  \undefined \def \bjtitle#1{#1}\fi
\ifx \bvolume  \undefined \def \bvolume#1{\textbf{#1}}\fi
\ifx \byear  \undefined \def \byear#1{#1}\fi
\ifx \bissue  \undefined \def \bissue#1{#1}\fi
\ifx \bfpage  \undefined \def \bfpage#1{#1}\fi
\ifx \blpage  \undefined \def \blpage #1{#1}\fi
\ifx \burl  \undefined \def \burl#1{\textsf{#1}}\fi
\ifx \doiurl  \undefined \def \doiurl#1{\url{https://doi.org/#1}}\fi
\ifx \betal  \undefined \def \betal{\textit{et al.}}\fi
\ifx \binstitute  \undefined \def \binstitute#1{#1}\fi
\ifx \binstitutionaled  \undefined \def \binstitutionaled#1{#1}\fi
\ifx \bctitle  \undefined \def \bctitle#1{#1}\fi
\ifx \beditor  \undefined \def \beditor#1{#1}\fi
\ifx \bpublisher  \undefined \def \bpublisher#1{#1}\fi
\ifx \bbtitle  \undefined \def \bbtitle#1{#1}\fi
\ifx \bedition  \undefined \def \bedition#1{#1}\fi
\ifx \bseriesno  \undefined \def \bseriesno#1{#1}\fi
\ifx \blocation  \undefined \def \blocation#1{#1}\fi
\ifx \bsertitle  \undefined \def \bsertitle#1{#1}\fi
\ifx \bsnm \undefined \def \bsnm#1{#1}\fi
\ifx \bsuffix \undefined \def \bsuffix#1{#1}\fi
\ifx \bparticle \undefined \def \bparticle#1{#1}\fi
\ifx \barticle \undefined \def \barticle#1{#1}\fi
\bibcommenthead
\ifx \bconfdate \undefined \def \bconfdate #1{#1}\fi
\ifx \botherref \undefined \def \botherref #1{#1}\fi
\ifx \url \undefined \def \url#1{\textsf{#1}}\fi
\ifx \bchapter \undefined \def \bchapter#1{#1}\fi
\ifx \bbook \undefined \def \bbook#1{#1}\fi
\ifx \bcomment \undefined \def \bcomment#1{#1}\fi
\ifx \oauthor \undefined \def \oauthor#1{#1}\fi
\ifx \citeauthoryear \undefined \def \citeauthoryear#1{#1}\fi
\ifx \endbibitem  \undefined \def \endbibitem {}\fi
\ifx \bconflocation  \undefined \def \bconflocation#1{#1}\fi
\ifx \arxivurl  \undefined \def \arxivurl#1{\textsf{#1}}\fi
\csname PreBibitemsHook\endcsname

%%% 1
\bibitem[\protect\citeauthoryear{Mather}{1965}]{MATHER0}
\begin{barticle}
\bauthor{\bsnm{Mather}, \binits{J.W.}}:
\batitle{Formation of a high-density deuterium plasma focus}.
\bjtitle{Phys. Fluids}
\bvolume{8}(\bissue{2}),
\bfpage{366}
(\byear{1965})
\end{barticle}
\endbibitem

%%% 2
\bibitem[\protect\citeauthoryear{Mather}{1971}]{MATHER}
\begin{barticle}
\bauthor{\bsnm{Mather}, \binits{J.W.}}:
\batitle{Dense plasma focus}.
\bjtitle{Methods of Exp. Phys.}
\bvolume{9B},
\bfpage{187}--\blpage{249}
(\byear{1971})
\end{barticle}
\endbibitem

%%% 3
\bibitem[\protect\citeauthoryear{Scholz}{2014}]{SCHOLZ}
\begin{bbook}
\bauthor{\bsnm{Scholz}, \binits{M.}}:
\bbtitle{Plasma-Focus and Controlled Nuclear Fusion}.
\bpublisher{Institute of Nuclear Physics of the Polish Academy of Sciences
  (PAN)},
\blocation{Kraków}
(\byear{2014})
\end{bbook}
\endbibitem

%%% 4
\bibitem[\protect\citeauthoryear{Auluck
  et~al.}{2021}]{AULUCKSADOWSKIMIKLASZEWSKI}
\begin{barticle}
\bauthor{\bsnm{Auluck}, \binits{S.K.H.}},
\bauthor{\bsnm{Kubes}, \binits{P.}},
\bauthor{\bsnm{Paduch}, \binits{M.}},
\bauthor{\bsnm{Sadowski}, \binits{M.J.}},
\bauthor{\bsnm{Krauz}, \binits{V.I.}},
\bauthor{\bsnm{Lee}, \binits{S.}},
\bauthor{\bsnm{Soto}, \binits{L.}},
\bauthor{\bsnm{Scholz}, \binits{M.}},
\bauthor{\bsnm{Miklaszewski}, \binits{R.}},
\bauthor{\bsnm{Schmidt}, \binits{H.}}, \betal:
\batitle{Update on the scientific status of the plasma focus}.
\bjtitle{Plasma}
\bvolume{4},
\bfpage{450}--\blpage{669}
(\byear{2021})
\doiurl{10.3390/plasma4030033}
\end{barticle}
\endbibitem

%%% 5
\bibitem[\protect\citeauthoryear{Haines}{2011}]{AreviewofthedenseZpinchHAINES}
\begin{barticle}
\bauthor{\bsnm{Haines}, \binits{M.}}:
\batitle{A review of the dense {Z}-pinch}.
\bjtitle{Plasma Phys. Control. Fusion}
\bvolume{53},
\bfpage{093001}
(\byear{2011})
\doiurl{10.1088/0741-3335/53/9/093001}
\end{barticle}
\endbibitem

%%% 6
\bibitem[\protect\citeauthoryear{Pavez and Soto}{2010}]{PAVEZ}
\begin{barticle}
\bauthor{\bsnm{Pavez}, \binits{C.}},
\bauthor{\bsnm{Soto}, \binits{L.}}:
\batitle{Demonstration of {X}-ray emission from an ultraminiature pinch plasma
  focus discharge operating at 0.1 {J} nanofocus}.
\bjtitle{IEEE Trans. on Plasma Science}
\bvolume{38}(\bissue{5}),
\bfpage{1132}--\blpage{1135}
(\byear{2010})
\doiurl{10.1109/TPS.2010.2045110}
\end{barticle}
\endbibitem

%%% 7
\bibitem[\protect\citeauthoryear{Brzosko et~al.}{1997}]{BRZOSKO}
\begin{bchapter}
\bauthor{\bsnm{Brzosko}, \binits{J.S.}},
\bauthor{\bsnm{Degnan}, \binits{J.H.}},
\bauthor{\bsnm{Filippov}, \binits{N.V.}},
\bauthor{\bsnm{Freeman}, \binits{B.L.}},
\bauthor{\bsnm{Kiutlu}, \binits{G.F.}},
\bauthor{\bsnm{Mather}, \binits{J.W.}}:
\bctitle{Comments on the feasibility of achieving scientific break-even with a
  plasma focus machine}.
In: \beditor{\bsnm{Panarella}, \binits{E.}} (ed.)
\bbtitle{Current Trends in International Fusion Research},
pp. \bfpage{11}--\blpage{32}.
\bpublisher{Springer},
\blocation{Boston, USA}
(\byear{1997}).
\burl{https://link.springer.com/book/10.1007/978-1-4615-5867-5}
\end{bchapter}
\endbibitem

%%% 8
\bibitem[\protect\citeauthoryear{Rager}{1982}]{RAGER}
\begin{bchapter}
\bauthor{\bsnm{Rager}, \binits{J.P.}}:
\bctitle{The plasma focus}.
In: \beditor{\bsnm{Brunelli}, \binits{B.}}, \betal (eds.)
\bbtitle{Unconventional Approaches to Fusion}.
\bpublisher{Plenum},
\blocation{New York}
(\byear{1982})
\end{bchapter}
\endbibitem

%%% 9
\bibitem[\protect\citeauthoryear{Soto}{2005}]{PPCFSOTO}
\begin{barticle}
\bauthor{\bsnm{Soto}, \binits{L.}}:
\batitle{New trends and future perspectives on plasma focus research}.
\bjtitle{Plasma Phys. Control. Fusion}
\bvolume{47},
\bfpage{361}--\blpage{381}
(\byear{2005})
\doiurl{10.1088/0741-3335/47/5A/027}
\end{barticle}
\endbibitem

%%% 10
\bibitem[\protect\citeauthoryear{Lee}{2009}]{LEE}
\begin{barticle}
\bauthor{\bsnm{Lee}, \binits{S.}}:
\batitle{Neutron yield saturation in plasma focus: A fundamental cause}.
\bjtitle{Applied Phys. Lett.}
\bvolume{95},
\bfpage{151503}
(\byear{2009})
\end{barticle}
\endbibitem

%%% 11
\bibitem[\protect\citeauthoryear{Damideh et~al.}{2025}]{DAMIDEH}
\begin{barticle}
\bauthor{\bsnm{Damideh}, \binits{V.}}, \betal:
\batitle{Experimental results and analysis of plasma dynamics and radiation
  output of the 100 k{V} dense plasma focus {FAETON-I}}.
\bjtitle{Nature Scientific Reports}
\bvolume{15},
\bfpage{23048}
(\byear{2025})
\doiurl{10.1038/s41598-025-07939-x}
\end{barticle}
\endbibitem

%%% 12
\bibitem[\protect\citeauthoryear{Nukulin and Polukhin}{2007}]{NUKULIN}
\begin{barticle}
\bauthor{\bsnm{Nukulin}, \binits{V.Y.}},
\bauthor{\bsnm{Polukhin}, \binits{S.N.}}:
\batitle{Saturation of the neutron yield from megajoule plasma focus
  facilities}.
\bjtitle{Plasma Physics Reports}
\bvolume{33}(\bissue{4}),
\bfpage{271}--\blpage{277}
(\byear{2007})
\end{barticle}
\endbibitem

%%% 13
\bibitem[\protect\citeauthoryear{Angeli et~al.}{2005}]{ANGELI}
\begin{barticle}
\bauthor{\bsnm{Angeli}, \binits{E.}}, \betal:
\batitle{Production of radioisotopes within a plasma focus device}.
\bjtitle{Nuclear Technology and Radiation Protection}
\bvolume{1},
\bfpage{33}--\blpage{37}
(\byear{2005})
\end{barticle}
\endbibitem

%%% 14
\bibitem[\protect\citeauthoryear{Lawson}{1957}]{LAWSON}
\begin{barticle}
\bauthor{\bsnm{Lawson}, \binits{J.D.}}:
\batitle{Some criteria for a power producing thermonuclear reactor}.
\bjtitle{Proceedings of the Physical Society. Section B}
\bvolume{70},
\bfpage{6}--\blpage{10}
(\byear{1957})
\doiurl{10.1088/0370-1301/70/1/303}
\end{barticle}
\endbibitem

%%% 15
\bibitem[\protect\citeauthoryear{Sadowski et~al.}{2020}]{SADOWSKI}
\begin{barticle}
\bauthor{\bsnm{Sadowski}, \binits{M.J.}}, \betal:
\batitle{Recent achievements of plasma studies within {PF-1000U} facility}.
\bjtitle{Acta Physica Polonica A}
\bvolume{138}(\bissue{4}),
\bfpage{613}--\blpage{621}
(\byear{2020})
\doiurl{10.12693/APhysPolA.138.613}
\end{barticle}
\endbibitem

%%% 16
\bibitem[\protect\citeauthoryear{Zucker et~al.}{1977}]{ZUCKER}
\begin{barticle}
\bauthor{\bsnm{Zucker}, \binits{O.}},
\bauthor{\bsnm{Bostick}, \binits{W.}},
\bauthor{\bsnm{Long}, \binits{J.}},
\bauthor{\bsnm{Luce}, \binits{J.}},
\bauthor{\bsnm{Sahlin}, \binits{H.}}:
\batitle{The plasma focus as a large fluence neutron source}.
\bjtitle{Nucl. Instrum. Methods}
\bvolume{145},
\bfpage{85}--\blpage{190}
(\byear{1977})
\end{barticle}
\endbibitem

%%% 17
\bibitem[\protect\citeauthoryear{Herold et~al.}{1999}]{HEROLD}
\begin{barticle}
\bauthor{\bsnm{Herold}, \binits{H.}},
\bauthor{\bsnm{Jerzykiewicz}, \binits{A.}},
\bauthor{\bsnm{Sadowski}, \binits{M.}},
\bauthor{\bsnm{Schmidt}, \binits{H.}}:
\batitle{Comparative analysis of large plasma focus experiments performed at
  {IPF}, stuttgart, and at {IPJ}, {\'s}wierk}.
\bjtitle{Nucl. Fusion}
\bvolume{29}(\bissue{8}),
\bfpage{1255}--\blpage{1269}
(\byear{1999})
\end{barticle}
\endbibitem

%%% 18
\bibitem[\protect\citeauthoryear{Auluck}{2023}]{AULUCKSATURATION}
\begin{barticle}
\bauthor{\bsnm{Auluck}, \binits{S.K.H.}}:
\batitle{On the failure of neutron yield scaling in the dense plasma focus}.
\bjtitle{Physics of Plasmas}
\bvolume{30},
\bfpage{080701}
(\byear{2023})
\doiurl{10.1063/5.0157626}
\end{barticle}
\endbibitem

%%% 19
\bibitem[\protect\citeauthoryear{Krishnan}{2012}]{KRISHNAN}
\begin{barticle}
\bauthor{\bsnm{Krishnan}, \binits{M.}}:
\batitle{The dense plasma focus: A versatile dense pinch for diverse
  applications}.
\bjtitle{IEEE Trans. on Plasma Science}
\bvolume{40}(\bissue{12}),
\bfpage{3189}--\blpage{3221}
(\byear{2012})
\doiurl{10.1109/TPS.2012.2222676}
\end{barticle}
\endbibitem

%%% 20
\bibitem[\protect\citeauthoryear{Hardy}{2004}]{HARDY}
\begin{botherref}
\oauthor{\bsnm{Hardy}, \binits{R.L.}}:
The plasma focus as a thruster.
Master's thesis,
Texas A\&M University
(2004).
\url{https://oaktrust.library.tamu.edu/server/api/core/bitstreams/c50dbc44-b999-4490-a6a6-06bab2ada2af/content}
\end{botherref}
\endbibitem

%%% 21
\bibitem[\protect\citeauthoryear{Thomas et~al.}{2005}]{THOMAS}
\begin{bchapter}
\bauthor{\bsnm{Thomas}, \binits{R.}}, \betal:
\bctitle{Advancements in dense plasma focus ({DPF}) for space propulsion}.
In: \beditor{\bsnm{El-Genk}, \binits{M.S.}} (ed.)
\bbtitle{Space Technology Applications International Forum 2005}.
\bsertitle{American Institute of Physics Conf. Proc.},
vol. \bseriesno{746},
pp. \bfpage{536}--\blpage{543}
(\byear{2005}).
\doiurl{10.1063/1.1867170}
\end{bchapter}
\endbibitem

%%% 22
\bibitem[\protect\citeauthoryear{Housley et~al.}{2021}]{HOUSLEY}
\begin{botherref}
\oauthor{\bsnm{Housley}, \binits{D.}}, et al.:
Effect of insulator surface conditioning on the pinch dynamics and {X}-ray
  production of a {Ne} filled dense plasma focus.
J. Applied Phys.
\textbf{129}(22)
(2021)
\end{botherref}
\endbibitem

%%% 23
\bibitem[\protect\citeauthoryear{Hahn}{2022}]{HAHN}
\begin{barticle}
\bauthor{\bsnm{Hahn}, \binits{E.N.}}:
\batitle{Effect of insulator length and fill pressure on filamentation and
  neutron production in a 4.6 k{V} dense plasma focus}.
\bjtitle{Phys. Plasmas}
\bvolume{29},
\bfpage{083508}
(\byear{2022})
\doiurl{10.1063/5.0087901}
\end{barticle}
\endbibitem

%%% 24
\bibitem[\protect\citeauthoryear{Nardi et~al.}{1988a}]{NARDI}
\begin{barticle}
\bauthor{\bsnm{Nardi}, \binits{V.}},
\bauthor{\bsnm{Bortolotti}, \binits{A.}},
\bauthor{\bsnm{Brzosko}, \binits{J.S.}},
\bauthor{\bsnm{Esper}, \binits{M.}},
\bauthor{\bsnm{Luo}, \binits{C.M.}},
\bauthor{\bsnm{Pedrielli}, \binits{F.}},
\bauthor{\bsnm{Powell}, \binits{C.}},
\bauthor{\bsnm{Zeng}, \binits{D.}}:
\batitle{Stimulated acceleration and confinement of deuterons in focused
  discharges - part {I}}.
\bjtitle{IEEE Trans. on Plasma Sci.}
\bvolume{16}(\bissue{3}),
\bfpage{368}
(\byear{1988})
\end{barticle}
\endbibitem

%%% 25
\bibitem[\protect\citeauthoryear{Nardi et~al.}{1988b}]{NARDI2}
\begin{barticle}
\bauthor{\bsnm{Nardi}, \binits{V.}},
\bauthor{\bsnm{Bilbao}, \binits{L.}},
\bauthor{\bsnm{Brzosko}, \binits{J.S.}},
\bauthor{\bsnm{Powell}, \binits{C.}},
\bauthor{\bsnm{Zeng}, \binits{D.}},
\bauthor{\bsnm{Bortolotti}, \binits{A.}},
\bauthor{\bsnm{Mezzetti}, \binits{F.}},
\bauthor{\bsnm{Robouch}, \binits{V.}}:
\batitle{Stimulated acceleration and confinement of deuterons in focused
  discharges - part {II}}.
\bjtitle{Trans. on Plasma Sci.}
\bvolume{16}(\bissue{3}),
\bfpage{374}
(\byear{1988})
\end{barticle}
\endbibitem

%%% 26
\bibitem[\protect\citeauthoryear{Gratton and Vargas}{1983}]{GRATTONVARGAS}
\begin{bchapter}
\bauthor{\bsnm{Gratton}, \binits{F.}},
\bauthor{\bsnm{Vargas}, \binits{J.M.}}:
\bctitle{Two-dimensional electromechanical model of the plasma focus}.
In: \beditor{\bsnm{Nardi}, \binits{V.}},
\beditor{\bsnm{Sahlin}, \binits{H.}},
\beditor{\bsnm{Bostick}, \binits{W.H.}} (eds.)
\bbtitle{Energy Storage, Compression and Switching, Vol. II},
pp. \bfpage{353}--\blpage{386}.
\bpublisher{Plenum},
\blocation{New York}
(\byear{1983})
\end{bchapter}
\endbibitem

%%% 27
\bibitem[\protect\citeauthoryear{Goyon et~al.}{2025}]{GOYONMJOLNIR}
\begin{barticle}
\bauthor{\bsnm{Goyon}, \binits{C.}}, \betal:
\batitle{Neutron generation dynamics inside a {MA}-class dense plasma focus
  {Z}-pinch}.
\bjtitle{Phys. Plasmas}
\bvolume{32},
\bfpage{033105}
(\byear{2025})
\end{barticle}
\endbibitem

%%% 28
\bibitem[\protect\citeauthoryear{Lerner et~al.}{2017}]{LERNERCONFINEMENT}
\begin{barticle}
\bauthor{\bsnm{Lerner}, \binits{E.J.}},
\bauthor{\bsnm{Hassan}, \binits{S.M.}},
\bauthor{\bsnm{Karamitsos}, \binits{I.}},
\bauthor{\bsnm{Von~Roessel}, \binits{F.}}:
\batitle{Confined ion energy $>$200 kev and increased fusion yield in a {DPF}
  with monolithic tungsten electrodes and pre-ionization}.
\bjtitle{Phys. of Plasmas}
\bvolume{24},
\bfpage{102708}
(\byear{2017})
\doiurl{10.1063/1.4989859}
\end{barticle}
\endbibitem

%%% 29
\bibitem[\protect\citeauthoryear{Lerner and Yousefi}{2014}]{LERNERRUNAWAY}
\begin{barticle}
\bauthor{\bsnm{Lerner}, \binits{E.J.}},
\bauthor{\bsnm{Yousefi}, \binits{H.R.}}:
\batitle{Runaway electrons as a source of impurity and reduced fusion yield in
  the dense plasma focus}.
\bjtitle{Phys. Plasmas}
\bvolume{21},
\bfpage{102706}
(\byear{2014})
\doiurl{10.1063/1.4898733}
\end{barticle}
\endbibitem

%%% 30
\bibitem[\protect\citeauthoryear{Lerner et~al.}{2023}]{LERNER}
\begin{botherref}
\oauthor{\bsnm{Lerner}, \binits{E.J.}}, et al.:
Focus fusion: Overview of progress towards p-$^{11}${B} fusion with the dense
  plasma focus.
J. Fusion Energy
\textbf{42}(7)
(2023)
\doiurl{10.1007/s10894-023-00345-z}
\end{botherref}
\endbibitem

%%% 31
\bibitem[\protect\citeauthoryear{Bostick et~al.}{1975}]{BOSTICK}
\begin{barticle}
\bauthor{\bsnm{Bostick}, \binits{W.H.}},
\bauthor{\bsnm{Nardi}, \binits{V.}},
\bauthor{\bsnm{Prior}, \binits{W.}}:
\batitle{Production and confinement of high-density plasmas}.
\bjtitle{Ann. N.Y. Acad. Sci.}
\bvolume{251},
\bfpage{2}--\blpage{29}
(\byear{1975})
\end{barticle}
\endbibitem

%%% 32
\bibitem[\protect\citeauthoryear{Haines}{2006}]{HAINES2006}
\begin{barticle}
\bauthor{\bsnm{Haines}, \binits{M.}}:
\batitle{Ion viscous heating in a magnetohydrodynamically unstable {Z} pinch at
  over $2 \cdot 10^9$ kelvin}.
\bjtitle{Phys. Rev. Lett.}
\bvolume{96},
\bfpage{075003}
(\byear{2006})
\end{barticle}
\endbibitem

%%% 33
\bibitem[\protect\citeauthoryear{Baalrud}{2025}]{BAALRUD}
\begin{barticle}
\bauthor{\bsnm{Baalrud}, \binits{S.D.}}:
\batitle{Constraints of radiation loss on thermonuclear reactor designs based
  on burning p-$^{11}${B} in a dense plasma focus}.
\bjtitle{Phys. Plasmas}
\bvolume{32},
\bfpage{102709}
(\byear{2025})
\doiurl{10.1063/5.0292235}
\end{barticle}
\endbibitem

%%% 34
\bibitem[\protect\citeauthoryear{Di~Vita}{2013}]{DIVITAPB}
\begin{barticle}
\bauthor{\bsnm{Di~Vita}, \binits{A.}}:
\batitle{On some necessary conditions for p-$^{11}${B} ignition in the hot
  spots of a plasma focus}.
\bjtitle{Eur. Phys. J. D}
\bvolume{67},
\bfpage{191}
(\byear{2013})
\doiurl{10.1140/epjd/e2013-40096-3}
\end{barticle}
\endbibitem

%%% 35
\bibitem[\protect\citeauthoryear{Lee}{2014}]{LEE2014}
\begin{barticle}
\bauthor{\bsnm{Lee}, \binits{S.}}:
\batitle{Plasma focus radiative model: Review of the {Lee} model code}.
\bjtitle{J. Fusion Energy}
\bvolume{33},
\bfpage{319}--\blpage{335}
(\byear{2014})
\doiurl{10.1007/s10894-014-9683-8}
\end{barticle}
\endbibitem

%%% 36
\bibitem[\protect\citeauthoryear{Scholz et~al.}{2004}]{N214}
\begin{barticle}
\bauthor{\bsnm{Scholz}, \binits{M.}}, \betal:
\batitle{The physics of a plasma focus}.
\bjtitle{Czechoslovak Journal of Physics}
\bvolume{54}(\bissue{Suppl. C}),
\bfpage{170}--\blpage{185}
(\byear{2004})
\end{barticle}
\endbibitem

%%% 37
\bibitem[\protect\citeauthoryear{Di~Vita}{1993}]{DIVITA1993}
\begin{barticle}
\bauthor{\bsnm{Di~Vita}, \binits{A.}}:
\batitle{The filamentary structure in the accelerating plasma sheath of a
  plasma focus: a simplified three-dimensional analysis}.
\bjtitle{J. Plasma Physics}
\bvolume{50}(\bissue{1}),
\bfpage{1}--\blpage{19}
(\byear{1993})
\end{barticle}
\endbibitem

%%% 38
\bibitem[\protect\citeauthoryear{Auluck}{2022}]{AULUCKFILAMENTATION}
\begin{barticle}
\bauthor{\bsnm{Auluck}, \binits{S.K.H.}}:
\batitle{On filamentation in the dense plasma focus}.
\bjtitle{Phys. Plasmas}
\bvolume{29},
\bfpage{030703}
(\byear{2022})
\doiurl{10.1063/5.0085870}
\end{barticle}
\endbibitem

%%% 39
\bibitem[\protect\citeauthoryear{Kwek et~al.}{1990}]{KWEK}
\begin{barticle}
\bauthor{\bsnm{Kwek}, \binits{K.H.}},
\bauthor{\bsnm{Tou}, \binits{T.Y.}},
\bauthor{\bsnm{Lee}, \binits{S.}}:
\batitle{Current sheath structure of the plasma focus in the run-down phase}.
\bjtitle{IEEE Trans. on Plasma Science}
\bvolume{18}(\bissue{5}),
\bfpage{826}--\blpage{830}
(\byear{1990})
\end{barticle}
\endbibitem

%%% 40
\bibitem[\protect\citeauthoryear{Auluck}{2021}]{AULUCKREPRESENTATION}
\begin{barticle}
\bauthor{\bsnm{Auluck}, \binits{S.K.H.}}:
\batitle{On the representation of dense plasma focus as a circuit element}.
\bjtitle{Phys. Plasmas}
\bvolume{28},
\bfpage{030703}
(\byear{2021})
\end{barticle}
\endbibitem

%%% 41
\bibitem[\protect\citeauthoryear{Lee and Saw}{2010}]{LEESAW}
\begin{botherref}
\oauthor{\bsnm{Lee}, \binits{S.}},
\oauthor{\bsnm{Saw}, \binits{S.H.}}:
Introduction to the Plasma Focus.
Joint ICTP-IAEA Workshop on Dense Magnetized Plasma and Plasma Diagnostics, 15
  - 26 November 2010 Trieste, Italy
(2010)
\end{botherref}
\endbibitem

%%% 42
\bibitem[\protect\citeauthoryear{Lee}{2008}]{LEE2008}
\begin{barticle}
\bauthor{\bsnm{Lee}, \binits{S.}}:
\batitle{Current and neutron scaling for megajoule plasma focus machines}.
\bjtitle{Plasma Phys. Control. Fusion}
\bvolume{50},
\bfpage{105005}
(\byear{2008})
\end{barticle}
\endbibitem

%%% 43
\bibitem[\protect\citeauthoryear{Pavez et~al.}{2023}]{ZORONDO}
\begin{barticle}
\bauthor{\bsnm{Pavez}, \binits{C.}}, \betal:
\batitle{New evidence about the nature of plasma filaments in plasma
  accelerators of type plasma-focus}.
\bjtitle{Plasma Phys. Control. Fusion}
\bvolume{65},
\bfpage{015003}
(\byear{2023})
\doiurl{10.1088/1361-6587/aca358}
\end{barticle}
\endbibitem

%%% 44
\bibitem[\protect\citeauthoryear{Auluck}{2024}]{AULUCKPOLOIDAL}
\begin{barticle}
\bauthor{\bsnm{Auluck}, \binits{S.K.H.}}:
\batitle{Poloidal magnetic field in the dense plasma focus}.
\bjtitle{Phys. Plasmas}
\bvolume{31},
\bfpage{010704}
(\byear{2024})
\end{barticle}
\endbibitem

%%% 45
\bibitem[\protect\citeauthoryear{Potter}{1971}]{POTTER}
\begin{barticle}
\bauthor{\bsnm{Potter}, \binits{D.E.}}:
\batitle{Numerical studies of the plasma focus}.
\bjtitle{Phys. Fluids}
\bvolume{14}(\bissue{9}),
\bfpage{1911}--\blpage{1923}
(\byear{1971})
\end{barticle}
\endbibitem

%%% 46
\bibitem[\protect\citeauthoryear{Soto et~al.}{2014}]{SOTO}
\begin{barticle}
\bauthor{\bsnm{Soto}, \binits{L.}},
\bauthor{\bsnm{Pavez}, \binits{C.}},
\bauthor{\bsnm{Castillo}, \binits{F.}},
\bauthor{\bsnm{Veloso}, \binits{F.}},
\bauthor{\bsnm{Moreno}, \binits{J.}},
\bauthor{\bsnm{Auluck}, \binits{S.K.H.}}:
\batitle{Filamentary structures in dense plasma focus: Current filaments or
  vortex filaments?}
\bjtitle{Phys. Plasmas}
\bvolume{21},
\bfpage{72702}
(\byear{2014})
\end{barticle}
\endbibitem

%%% 47
\bibitem[\protect\citeauthoryear{Auluck}{2009}]{AULUCKDESCRIPTION}
\begin{barticle}
\bauthor{\bsnm{Auluck}, \binits{S.K.H.}}:
\batitle{Description of plasma focus current sheath as the {Turner} relaxed
  state of a {Hall} magnetofluid}.
\bjtitle{Physics of Plasmas}
\bvolume{16},
\bfpage{122504}
(\byear{2009})
\doiurl{10.1063/1.3270112}
\end{barticle}
\endbibitem

%%% 48
\bibitem[\protect\citeauthoryear{Bergmans et~al.}{1999}]{BERGMANS}
\begin{barticle}
\bauthor{\bsnm{Bergmans}, \binits{J.}},
\bauthor{\bsnm{Kuvshinov}, \binits{B.N.}},
\bauthor{\bsnm{Lakhin}, \binits{V.P.}},
\bauthor{\bsnm{Schep}, \binits{T.J.}},
\bauthor{\bsnm{Westerhof}, \binits{E.}}:
\batitle{Current-vortex filaments in magnetized plasmas}.
\bjtitle{Plasma Phys. Control. Fusion}
\bvolume{41},
\bfpage{709}--\blpage{717}
(\byear{1999})
\doiurl{10.1088/0741-3335/41/3A/064}
\end{barticle}
\endbibitem

%%% 49
\bibitem[\protect\citeauthoryear{Bostick}{1985}]{BOSTICKORIGINAL}
\begin{barticle}
\bauthor{\bsnm{Bostick}, \binits{W.H.}}:
\batitle{The morphology of the electron}.
\bjtitle{International J. Fusion Energy}
\bvolume{3}(\bissue{1}),
\bfpage{9}--\blpage{52}
(\byear{1985})
\end{barticle}
\endbibitem

%%% 50
\bibitem[\protect\citeauthoryear{Petviashvili}{1993}]{PETVIASHVILI}
\begin{bchapter}
\bauthor{\bsnm{Petviashvili}, \binits{V.I.}}:
\bctitle{Dynamics of vortex-current filaments in {MHD} plasma}.
In: \beditor{\bsnm{Fokas}, \binits{A.S.}}, \betal (eds.)
\bbtitle{Nonlinear Processes in Physics}.
\bsertitle{Springer Series in Nonlinear Dynamics}.
\bpublisher{Springer},
\blocation{Berlin Heidelberg}
(\byear{1993}).
\doiurl{10.1007/978-3-642-77769-1\_58} .
\burl{https://doi.org}
\end{bchapter}
\endbibitem

%%% 51
\bibitem[\protect\citeauthoryear{Bostick}{1987}]{WHBOSTICK}
\begin{barticle}
\bauthor{\bsnm{Bostick}, \binits{W.H.}}:
\batitle{Similarities between the plasma vortex filaments (relativistic and
  nonrelativistic) observed in the plasma focus and in conventional
  relativistic electron-beam machines}.
\bjtitle{Fusion Technology}
\bvolume{12}(\bissue{1}),
\bfpage{92}--\blpage{103}
(\byear{1987})
\end{barticle}
\endbibitem

%%% 52
\bibitem[\protect\citeauthoryear{Sestero et~al.}{1980}]{SESTERO}
\begin{barticle}
\bauthor{\bsnm{Sestero}, \binits{A.}},
\bauthor{\bsnm{Robouch}, \binits{B.V.}},
\bauthor{\bsnm{Podda}, \binits{S.}}:
\batitle{Suggested relaxation of plasma focus discharges to helical force-free
  configurations}.
\bjtitle{Plasma Phys.}
\bvolume{22},
\bfpage{1039}--\blpage{1041}
(\byear{1980})
\end{barticle}
\endbibitem

%%% 53
\bibitem[\protect\citeauthoryear{Di~Vita}{2009}]{DIVITAHOTSPOT}
\begin{barticle}
\bauthor{\bsnm{Di~Vita}, \binits{A.}}:
\batitle{Hot spots and filaments in the pinch of a plasma focus: a unified
  approach}.
\bjtitle{Eur. Phys. J. D}
\bvolume{54},
\bfpage{451}--\blpage{461}
(\byear{2009})
\doiurl{10.1140/epjd/e2009-00092-x}
\end{barticle}
\endbibitem

%%% 54
\bibitem[\protect\citeauthoryear{Casanova et~al.}{2012}]{CASANOVA}
\begin{barticle}
\bauthor{\bsnm{Casanova}, \binits{F.}}, \betal:
\batitle{Toroidal high-density singularity in a small plasma focus}.
\bjtitle{J. Fusion Energy}
\bvolume{31},
\bfpage{279}--\blpage{283}
(\byear{2012})
\doiurl{10.1007/s10894-011-9469-1}
\end{barticle}
\endbibitem

%%% 55
\bibitem[\protect\citeauthoryear{Turner}{1986}]{TURNER}
\begin{barticle}
\bauthor{\bsnm{Turner}, \binits{L.}}:
\batitle{{Hall} effects on magnetic relaxation}.
\bjtitle{IEEE Trans. on Plasma Science}
\bvolume{14}(\bissue{6}),
\bfpage{849}--\blpage{857}
(\byear{1986})
\doiurl{10.1109/TPS.1986.4316633}
\end{barticle}
\endbibitem

%%% 56
\bibitem[\protect\citeauthoryear{Di~Vita}{2010}]{DIVITAHARTMANN}
\begin{barticle}
\bauthor{\bsnm{Di~Vita}, \binits{A.}}:
\batitle{A lower bound on {Hartmann} number for relaxed plasmas described by
  {Taylor's} principle}.
\bjtitle{Eur. Phys. J. D}
\bvolume{56},
\bfpage{205}--\blpage{208}
(\byear{2010})
\doiurl{10.1140/epjd/e2009-00294-2}
\end{barticle}
\endbibitem

%%% 57
\bibitem[\protect\citeauthoryear{Nikulin et~al.}{2017}]{TSYBENKONIKULIN}
\begin{barticle}
\bauthor{\bsnm{Nikulin}, \binits{V.Y.}}, \betal:
\batitle{Supersonic, subsonic and stationary filaments in the plasma focus}.
\bjtitle{J. Phys.: Conf. Ser.}
\bvolume{907},
\bfpage{012024}
(\byear{2017})
\end{barticle}
\endbibitem

%%% 58
\bibitem[\protect\citeauthoryear{Nikulin et~al.}{2020}]{TSYBENKOERISKIN}
\begin{barticle}
\bauthor{\bsnm{Nikulin}, \binits{V.Y.}}, \betal:
\batitle{Structures and generation of current filaments in plasma focus}.
\bjtitle{Acta Physica Polonica A}
\bvolume{138}(\bissue{4}),
\bfpage{622}--\blpage{625}
(\byear{2020})
\end{barticle}
\endbibitem

%%% 59
\bibitem[\protect\citeauthoryear{Di~Vita
  et~al.}{2000}]{SPONTANEOUSSYMMETRYBREAKING}
\begin{bchapter}
\bauthor{\bsnm{Di~Vita}, \binits{A.}},
\bauthor{\bsnm{Paganucci}, \binits{F.}},
\bauthor{\bsnm{Rossetti}, \binits{P.}},
\bauthor{\bsnm{Andrenucci}, \binits{M.}}:
\bctitle{Spontaneous symmetry breaking in {MPD} plasmas}.
In: \bbtitle{AIAA/ASME/SAE/ASEE Joint Propulsion Conference and Exhibit, 36th},
\bconflocation{Huntsville, AL, USA}
(\byear{2000})
\end{bchapter}
\endbibitem

%%% 60
\bibitem[\protect\citeauthoryear{Lee and Serban}{1996}]{LEESERBAN}
\begin{barticle}
\bauthor{\bsnm{Lee}, \binits{S.}},
\bauthor{\bsnm{Serban}, \binits{A.}}:
\batitle{Dimensions and lifetime of the plasma focus pinch}.
\bjtitle{IEEE Trans. Plasma Science}
\bvolume{24}(\bissue{3}),
\bfpage{1101}--\blpage{1105}
(\byear{1996})
\end{barticle}
\endbibitem

%%% 61
\bibitem[\protect\citeauthoryear{Ferro~Fontan and
  Sicardi~Schifino}{1983}]{FERROFONTAN}
\begin{bchapter}
\bauthor{\bsnm{Ferro~Fontan}, \binits{C.}},
\bauthor{\bsnm{Sicardi~Schifino}, \binits{A.}}:
\bctitle{Generation of kilogauss radial magnetic fields in the plasma focus
  current sheath}.
In: \beditor{\bsnm{Nardi}, \binits{V.}},
\beditor{\bsnm{Sahlin}, \binits{H.}},
\beditor{\bsnm{Bostick}, \binits{W.H.}} (eds.)
\bbtitle{Energy Storage, Compression and Switching, Vol. 2},
p. \bfpage{607}.
\bpublisher{Plenum},
\blocation{New York, NY, USA}
(\byear{1983})
\end{bchapter}
\endbibitem

%%% 62
\bibitem[\protect\citeauthoryear{Beresnyak}{2023}]{NRL}
\begin{bbook}
\beditor{\bsnm{Beresnyak}, \binits{A.}} (ed.):
\bbtitle{2023 {NRL} Plasma Formulary}.
\bpublisher{U.S. Office of Naval Research}, \blocation{???}
(\byear{2023})
\end{bbook}
\endbibitem

%%% 63
\bibitem[\protect\citeauthoryear{Fruchtman and Strauss}{1992}]{FRUCHTMAN}
\begin{barticle}
\bauthor{\bsnm{Fruchtman}, \binits{A.}},
\bauthor{\bsnm{Strauss}, \binits{H.R.}}:
\batitle{Thermomagnetic instability in a magnetized plasma}.
\bjtitle{Phys. Fluids B}
\bvolume{4}(\bissue{6}),
\bfpage{1397}--\blpage{1400}
(\byear{1992})
\end{barticle}
\endbibitem

%%% 64
\bibitem[\protect\citeauthoryear{Stockem et~al.}{2008}]{STOCKEM}
\begin{bchapter}
\bauthor{\bsnm{Stockem}, \binits{A.}}, \betal:
\bctitle{Suppression of the filamentation instability by a flow-aligned
  magnetic field}.
In: \bbtitle{35th EPS Conference on Plasma Phys.}
\bsertitle{ECA Vol.},
vol. \bseriesno{32D},
pp. \bfpage{2}--\blpage{174}.
\bconflocation{Hersonissos}
(\byear{2008})
\end{bchapter}
\endbibitem

%%% 65
\bibitem[\protect\citeauthoryear{Bret and Perez~Alvaro}{2011}]{BRET}
\begin{botherref}
\oauthor{\bsnm{Bret}, \binits{A.}},
\oauthor{\bsnm{Perez~Alvaro}, \binits{E.}}:
Robustness of the filamentation instability as shock mediator in arbitrarily
  oriented magnetic field.
Physics of Plasmas
\textbf{18}(8)
(2011)
\end{botherref}
\endbibitem

%%% 66
\bibitem[\protect\citeauthoryear{Bud'ko et~al.}{1989}]{BUDKO}
\begin{barticle}
\bauthor{\bsnm{Bud'ko}, \binits{A.B.}}, \betal:
\batitle{Stability analysis of dynamic {Z} pinches and theta pinches}.
\bjtitle{Physics of Fluids B: Plasma Physics (1989-1993)}
\bvolume{1},
\bfpage{598}
(\byear{1989})
\doiurl{10.1063/1.859207}
\end{barticle}
\endbibitem

%%% 67
\bibitem[\protect\citeauthoryear{Mikitchuk et~al.}{2014}]{MIKITCHUK}
\begin{barticle}
\bauthor{\bsnm{Mikitchuk}, \binits{D.}}, \betal:
\batitle{Mitigation of instabilities in a {Z}-pinch plasma by a preembedded
  axial magnetic field}.
\bjtitle{IEEE Trans. on Plasma Science}
\bvolume{42}(\bissue{10}),
\bfpage{2524}--\blpage{2525}
(\byear{2014})
\end{barticle}
\endbibitem

%%% 68
\bibitem[\protect\citeauthoryear{Goebel and Katz}{2008}]{GOEBEL}
\begin{bbook}
\bauthor{\bsnm{Goebel}, \binits{D.M.}},
\bauthor{\bsnm{Katz}, \binits{I.}}:
\bbtitle{Fundamentals of Electric Propulsion: Ion and {Hall} Thrusters}.
\bsertitle{JPL Space Science and Technology Series}.
\bpublisher{John Wiley \& Sons},
\blocation{Hoboken, New Jersey}
(\byear{2008}).
\burl{https://nasa.gov}
\end{bbook}
\endbibitem

%%% 69
\bibitem[\protect\citeauthoryear{Mikhailov et~al.}{2026}]{MIKHAILOV}
\begin{barticle}
\bauthor{\bsnm{Mikhailov}, \binits{Y.V.}},
\bauthor{\bsnm{Prokuratova}, \binits{I.A.}},
\bauthor{\bsnm{Lemeshko}, \binits{B.D.}}:
\batitle{Physical processes in sealed plasma focus chambers that limit their
  service life}.
\bjtitle{Instruments and Experimental Techniques}
\bvolume{69}(\bissue{1}),
\bfpage{73}--\blpage{90}
(\byear{2026})
\end{barticle}
\endbibitem

%%% 70
\bibitem[\protect\citeauthoryear{Oreshkin and Oreshkin}{2021}]{ORESHKIN}
\begin{barticle}
\bauthor{\bsnm{Oreshkin}, \binits{V.I.}},
\bauthor{\bsnm{Oreshkin}, \binits{E.V.}}:
\batitle{Effect of the plasma self-radiation on the growth of thermal
  filamentation instabilities in imploding {Z} pinches}.
\bjtitle{Plasma Physics and Controlled Fusion}
\bvolume{63}(\bissue{12}),
\bfpage{125013}
(\byear{2021})
\end{barticle}
\endbibitem

%%% 71
\bibitem[\protect\citeauthoryear{Takeda et~al.}{2023}]{TAKEDA}
\begin{barticle}
\bauthor{\bsnm{Takeda}, \binits{S.}},
\bauthor{\bsnm{Keeley}, \binits{A.R.}},
\bauthor{\bsnm{Managi}, \binits{S.}}:
\batitle{How many years away is fusion energy? {A} review}.
\bjtitle{Journal of Fusion Energy}
\bvolume{42},
\bfpage{16}
(\byear{2023})
\end{barticle}
\endbibitem

%%% 72
\bibitem[\protect\citeauthoryear{Tang et~al.}{2026}]{TANGNOLL}
\begin{barticle}
\bauthor{\bsnm{Tang}, \binits{L.X.}},
\bauthor{\bsnm{Noll}, \binits{B.}},
\bauthor{\bsnm{Panda}, \binits{A.}},
\bauthor{\bsnm{Tobias}, \binits{S.S.}}:
\batitle{Fusion power experience rates are overestimated}.
\bjtitle{Nature Energy}
(\byear{2026})
\doiurl{10.1038/s41560-026-02023-8}
\end{barticle}
\endbibitem

%%% 73
\bibitem[\protect\citeauthoryear{Zoita and Lungu}{2001}]{ZOITA}
\begin{barticle}
\bauthor{\bsnm{Zoita}, \binits{V.}},
\bauthor{\bsnm{Lungu}, \binits{S.}}:
\batitle{A fusion-fission hybrid reactor driven by high-density pinch plasmas}.
\bjtitle{Nukleonika}
\bvolume{46}(\bissue{Supplement 1}),
\bfpage{81}--\blpage{84}
(\byear{2001})
\end{barticle}
\endbibitem

%%% 74
\bibitem[\protect\citeauthoryear{Soto et~al.}{2022}]{SOTOPRESENTATION}
\begin{botherref}
\oauthor{\bsnm{Soto}, \binits{L.}},
\oauthor{\bsnm{Clausse}, \binits{A.}},
\oauthor{\bsnm{Friedli}, \binits{C.}},
\oauthor{\bsnm{Altamirano}, \binits{L.}}:
The Concept of Plasma Focus driven Fusion-Fission Hybrid Reactors.
{IAEA} TM Synergies Between Nuclear Fusion Technology Developments and Advanced
  Nuclear Fission Technologies, Vienna, Austria
(2022).
\url{https://iaea.org}
\end{botherref}
\endbibitem

%%% 75
\bibitem[\protect\citeauthoryear{Sadat~Kiai et~al.}{2025}]{SADATKIAI}
\begin{barticle}
\bauthor{\bsnm{Sadat~Kiai}, \binits{S.M.}}, \betal:
\batitle{Double 3 {MJ} dense plasma focus for thermonuclear drive inertial
  confinement fusion}.
\bjtitle{Nature - Scientific Communications}
\bvolume{15},
\bfpage{14081}
(\byear{2025})
\doiurl{10.1038/s41598-025-96736-7}
\end{barticle}
\endbibitem

\end{thebibliography}
%% if required, the content of .bbl file can be included here once bbl is generated
%%\input sn-article.bbl

\end{document}